%% file: ANA-EXOT-2022-40-PAPER.tex
\documentclass[cernpreprint, PAPER, atlasdraft=false, texlive=2023, UKenglish, texmf, orcidlogo]{atlasdoc}
\usepackage{atlaspackage}
\usepackage{siunitx}
\usepackage{makecell}
\usepackage{atlasbiblatex}

\usepackage{atlasphysics}
\graphicspath{{logos/}{figures/}}

\usepackage{ANA-EXOT-2022-40-PAPER-defs}

\AtlasTitle{
Search for new particles in final states with a boosted top quark and missing transverse momentum in proton-proton collisions at $\sqrt{s}$=13 TeV with the ATLAS detector
}

\AtlasAbstract{%
A search for events with one top quark and missing transverse momentum in the final state is presented.
The fully hadronic decay of the top quark is explored by selecting events with a reconstructed boosted top-quark
topology produced in association with large missing transverse momentum.
The analysis uses $139$~fb$^{-1}$ of proton--proton collision data at a
centre-of-mass energy of  $\sqrt{s} =13$~\TeV recorded during 2015-2018
by the ATLAS detector at the Large Hadron Collider. The results are interpreted in the context of simplified models for Dark Matter particle
production and the single production of a vector-like $T$ quark.
Without significant excess relative to the Standard Model expectations,
95\% confidence-level upper limits on the corresponding cross-sections are obtained.
The production of Dark Matter particles in association with a single top quark is excluded
for masses of a scalar (vector) mediator up to $4.3$ ($2.3$)~\TeV,
assuming $m_\chi=1$~\GeV and the model couplings $\lambda_q=0.6$ and $\lambda_\chi=0.4$ ($a=0.5$ and $g_\chi=1$).
The production of a single vector-like $T$ quark is excluded for masses below $1.8$~\TeV assuming
a coupling to the top quark $\kappa_T=0.5$ and a branching ratio for $T\to Zt$ of 25\%.

}

\author{The ATLAS Collaboration}

\AtlasRefCode{EXOT-2022-40}

\PreprintIdNumber{CERN-EP-2024-046}

\AtlasJournalRef{JHEP 05 (2024) 263}
\AtlasDOI{10.1007/JHEP05(2024)263}

\hypersetup{pdftitle={ATLAS document},pdfauthor={The ATLAS Collaboration}}

\begin{document}

\maketitle

\section{Introduction}
\label{sec:intro}

Despite its success in predicting the phenomenology of fundamental particle physics, the Standard Model (SM) is still far from being considered a complete description of nature at the smallest length scales due to experimental observations such as the evidence for Dark Matter (DM)~\cite{annurev.aa.25.090187.002233, annurev-astro-082708-101659,BERTONE2005279} and theoretical arguments, one of these being the hierarchy problem in the Higgs sector~\cite{Susskind:1978ms}. The top quark, being the heaviest fundamental particle in the SM, plays an important role in the resolution of these questions in the context of many Beyond the Standard Model (BSM) extensions. This paper focuses on the \emph{mono-top} experimental signature and presents a search for an excess of events relative to the SM prediction in which a single top quark is produced together with significant missing transverse momentum ($\MET$). Data from proton--proton ($pp$) collisions at the Large Hadron Collider (LHC) corresponding to an integrated luminosity of $139$~fb$^{-1}$ at a centre-of-mass energy of $\sqrt{s}=13$~\TeV are used. The \emph{mono-top} signature can arise from the production of DM in association with a top quark, where the DM candidate is identified as missing transverse momentum in the detector~\cite{Abercrombie_2020,Andrea_2011,PhysRevD.96.035031,PANI20188}, or from the production of a single new vector-like $T$ quark~\cite{Aguilar_Saavedra_2009,Matsedonskyi_2014,Buchkremer_2013}, when $T$ decays into a top quark and a $Z$ boson, with the $Z$ boson decaying into neutrinos.

Strong evidence of non-luminous gravitating matter, commonly known as DM, arises from measurements of astrophysical phenomena such as the cosmic microwave background and gravitational lensing. Through its gravitational interactions, it is suggested that DM constitutes up to five times the ordinary matter contained in the universe~\cite{Hinshaw:2012aka,Akrami:2018vks}.
Its nature remains however unknown and is one of the major questions in physics. Assuming that DM is a weakly interacting massive particle~\cite{Steigman:1984ac}, these particles may be produced in $pp$ collisions at the LHC via new mediator particles that couple both to SM and DM particles. While such candidates are not expected to interact significantly with the detector, the momentum imbalance of SM particles produced in association with the unobserved DM particles could allow these processes to be detected.

The ATLAS and CMS Collaborations have carried out several searches for DM production in association with jets~\cite{EXOT-2018-06,CMS-EXO-20-004}, photons~\cite{EXOT-2018-63,CMS-EXO-16-053}, vector bosons~\cite{HIGG-2018-26,CMS-EXO-19-003,EXOT-2016-23,CMS-EXO-16-048}, Higgs bosons~\cite{EXOT-2018-46,HIGG-2019-02,HIGG-2019-02,CMS-EXO-18-011} and pairs of $b$-quarks~\cite{SUSY-2018-34, CMS-SUS-19-006} and top quarks~\cite{SUSY-2018-08,CMS-EXO-16-049,CMS-EXO-18-010,SUSY-2019-12,EXOT-2018-43,EXOT-2021-01}. Searches for same-charge top-quark pairs also probe the production of DM mediator particles~\cite{EXOT-2016-16}.

The \emph{mono-top}~signature targeted in this paper has already been explored by the CDF Collaboration in $p\pbar$ collisions at $\sqrt{s}=1.96$~\TeV~\cite{PhysRevLett.108.201802} and more recently by both the ATLAS and CMS Collaborations in $pp$ collisions at $\sqrt{s}=8$~\TeV~\cite{CMS-B2G-12-022,TOPQ-2013-11} and $\sqrt{s}=13$~\TeV~\cite{CMS-EXO-16-051,EXOT-2017-16}.
BSM models that predict a \emph{mono-top} signature are typically characterised by the violation of the baryon number or by the inclusion of flavour-changing neutral currents (FCNC)~\cite{Andrea_2011}. To consistently explore all the possible scenarios, an effective model is usually considered in which new mediators connect the SM particles and the DM candidates, encompassing all the tree-level production mechanisms of \emph{mono-top} events~\cite{Andrea_2011,Abercrombie_2020}.
Two mechanisms are expected depending on the nature of the mediator particle.
One case consists of the production of a new scalar mediator $\phi$ decaying into a top quark and a DM candidate $\chi$. The latest analyses exclude such production for masses of the mediator particle below $3.5$~\TeV. The second mechanism is characterised by the production of a vector mediator $V$ that couples with SM particles via an FCNC interaction and decays into DM particles $\chi$. Masses of the mediator $V$ below $2$~\TeV are excluded by the latest analyses targeting this model. This analysis focuses on the high mass region beyond the excluded phase space.

Vector-like quarks (VLQ) are colour-triplet spin-1/2 fermions in which the left- and right-handed components have the same properties under transformations of the $SU(2)_L \times U(1)_Y$ electroweak symmetry group. Such  particles are predicted in many BSM extensions, such as Little Higgs~\cite{Arkani_Hamed_2002,Schmaltz_2005} and Composite Higgs~\cite{KAPLAN1984187,AGASHE2005165} models, to cancel out the quadratically divergent contributions to the Higgs boson mass caused by radiative corrections from the SM particles. They are expected to mix with SM quarks and modify their Yukawa couplings depending on the specific VLQ model~\cite{DELAGUILA1983107,del_Aguila_2000}.
To preserve the gauge invariance, only a limited set of possible representations exist~\cite{del_Aguila_2000, Aguila_2000}. In this document, only the production of a single $2/3e$-charged vector-like $T$ quark in the  weak-isospin singlet is considered.
Even though the coupling of $T$ with quarks from the first two generations is not excluded~\cite{Atre_2009,  Atre_2011},  it is common to assume  $T$ quarks couple only with the third-generation quarks~\cite{Aguilar_Saavedra_2009}.

Vector-like quarks can be produced singly via the electroweak interaction or in pairs via the strong interaction in $pp$ collisions. While the cross-section for pair production is given by quantum chromodynamics, the single-production cross-section explicitly depends on the coupling of the VLQs to SM vector bosons.
There have been numerous searches for the pair production of VLQs~\cite{
CMS-EXO-11-005,
CMS-EXO-11-099,
EXOT-2011-22,
CMS-B2G-13-005,
CMS-B2G-13-006,
EXOT-2014-10,
EXOT-2013-18,
EXOT-2016-15,
CMS-B2G-16-024,
CMS-B2G-17-003,
CMS-B2G-17-011,
CMS-B2G-17-012,
EXOT-2016-14,
EXOT-2016-13,
EXOT-2017-34,
EXOT-2017-17,
EXOT-2018-58,
CMS-B2G-18-005,}\
that have excluded $T$-quark
masses below 1.37~\TeV at 95\%\ confidence level (CL) for a variety of decay modes.
For $T$-quark masses above  $\sim$1~\TeV, VLQs would mainly be produced singly
if the couplings to SM particles are sufficiently large.
Searches for single production of $T$~quarks have placed limits on $T$-quark
production cross-sections for $T$-quark masses between 1 and 2~\TeV
at 95\%\ CL for various couplings~\cite{
EXOT-2014-12,
EXOT-2014-13,
EXOT-2018-52,
EXOT-2020-01,
CMS-B2G-15-008,
CMS-B2G-16-001,
CMS-B2G-16-005,
CMS-B2G-16-013,
CMS-B2G-17-007,
CMS-B2G-17-018,
CMS-B2G-18-003}.
For these higher masses, where single VLQ production is expected to
dominate~\cite{Aguilar-Saavedra:2013qpa},
the cross-section depends on the VLQ mass as
well as the couplings to SM particles.
In this study, the coupling of $T$ quarks with top quarks is described by the generalised coupling $\kappa_T$~\cite{Buchkremer_2013,Matsedonskyi_2014}. The $T$ quarks can decay both via charged and neutral currents into  $Wb$, $Zt$, or $Ht$. A singlet $T$ with a branching ratio ${\mathcal{B}}(T\to Zt) = 25$\% is assumed. The $T\rightarrow Zt\rightarrow\nu\bar{\nu} q\bar{q}' b$ final state is the one considered in this search.

This paper follows a methodology similar to that used in the previous search based on the 36~fb$^{-1}$ $\sqrt{s}=13$~\TeV data sample~\cite{EXOT-2017-16}. It considers the channel in which the top quark decays hadronically, found to be the most sensitive in the previous analysis. Signal regions are defined to maximise the discovery potential, introducing an extreme gradient-boosted (\xgb) decision tree (BDT)~\cite{10.1145/2939672.2939785} to enhance the signal discrimination against the SM background. The output score of the \xgb algorithm is also used as the discriminating variable in the signal regions. Improvements in the reconstruction and calibration of the final-state particles  are also incorporated, in particular, related to jets and their flavour identification. Regions enriched in the main sources of background events are identified and referred to as control regions. A simultaneous fit to the data yields in the control regions and to the \xgb output discriminant distribution in the signal regions is performed to constrain the background prediction and determine the possible signal contribution. Systematic uncertainties are considered and incorporated as nuisance parameters in the fit. A set of validation regions is defined to assess the robustness of the extrapolation on the background prediction from the control regions to the signal regions.

This paper is organised as follows. The signal models are introduced in Section~\ref{sec:signal}. A brief description of the ATLAS detector is given in Section~\ref{sec:atlas}. The data and the simulated samples used in the analysis are described in Section~\ref{sec:samples}. The algorithms used for the reconstruction and identification of the final-state particles are summarised in Section~\ref{sec:reco}. The selection criteria for the events considered in the analysis and the control, validation and signal region definitions are described in Section~\ref{sec:selection}. An introduction to the \xgb classifier, used to separate the background from the signal and to define the signal regions, is also given in this section. The experimental and systematic uncertainties, summarised in Section~\ref{sec:systematics}, are accounted for in the statistical interpretation of data. The results are presented in Section~\ref{sec:results}. Conclusions are given in Section~\ref{sec:conclusions}.


%

%
\section{Signal phenomenology}
\label{sec:signal}

%

This section contains the description of the models considered in this paper. Details of the model predicting DM production in association with a single top quark are given in Section~\ref{subsec:dm}. The production of a singlet vector-like $T$ quark model is presented in Section~\ref{subsec:vlq}.

\subsection{Dark Matter production in association with top quarks}
\label{subsec:dm}
The DM production in association with top quarks is considered for a scalar or a vector DM mediator particle~\cite{Andrea_2011,Abercrombie_2020,PhysRevD.86.034008,FeynRulesMonotops}.

\begin{figure}[h]
\centering
\subfloat[]{\includegraphics[width=0.3\textwidth]{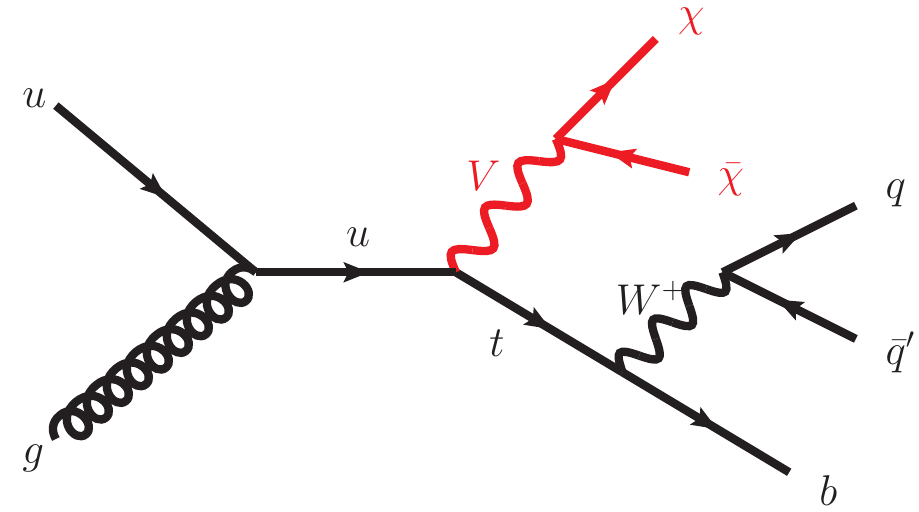}\label{img:feyn_nonres_1}} $\quad$
\subfloat[]{\includegraphics[width=0.3\textwidth]{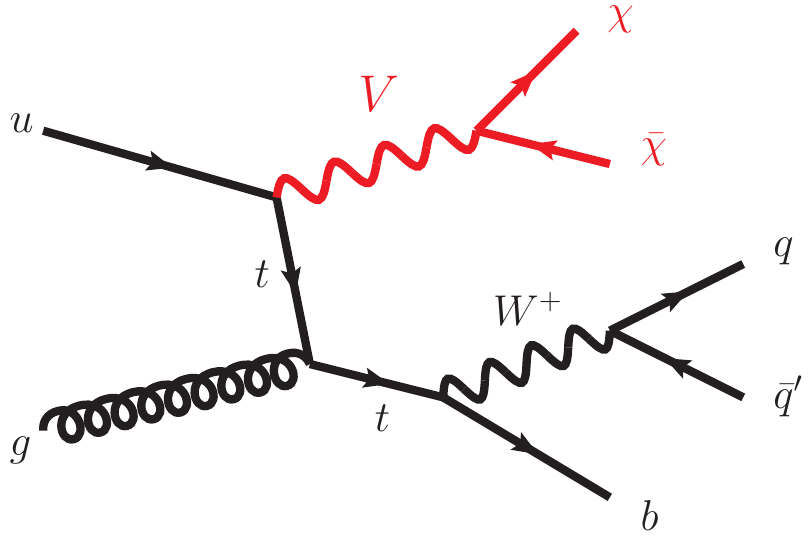}\label{img:feyn_nonres_2}} $\quad$
\subfloat[]{\includegraphics[width=0.33\textwidth]{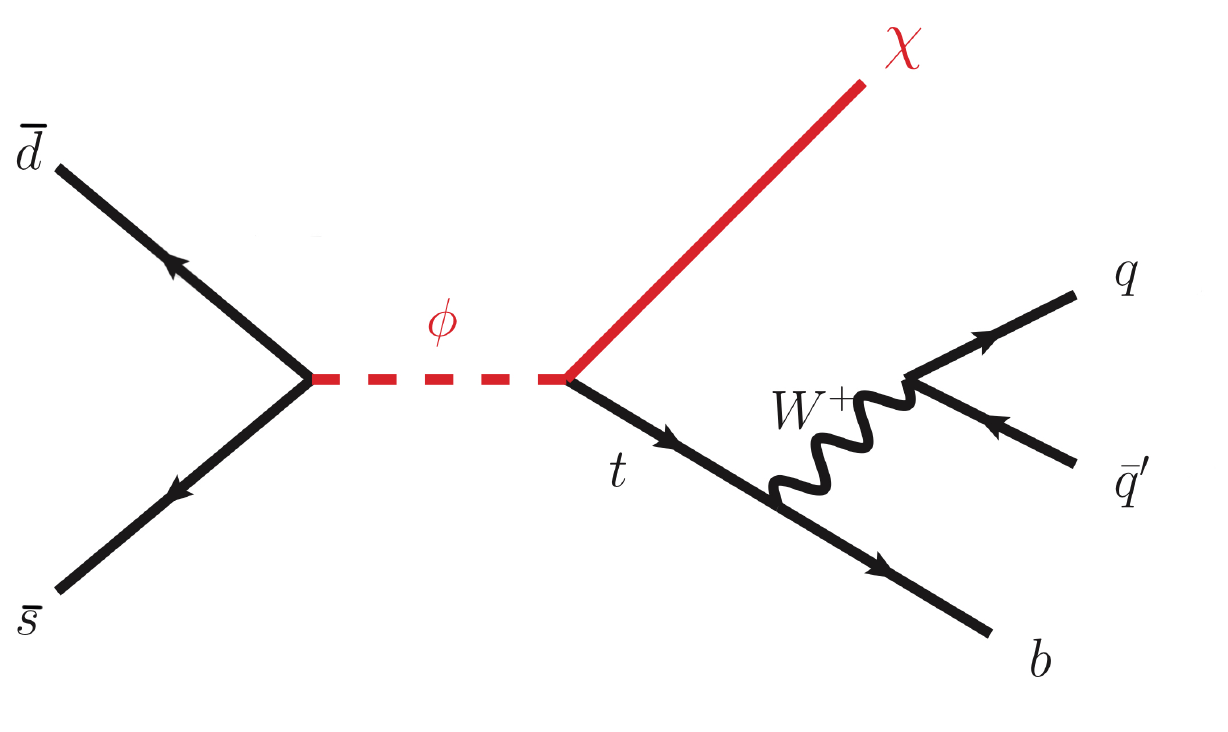}\label{img:feyn_res}}
\caption{Representative Feynman diagrams for the DM production in association with a single top-quark decaying hadronically in the (a,b) vector mediator (non-resonant) and (c) scalar mediator (resonant) cases. }
\label{fig:feynman-diagrams-dm}
\end{figure}

The vector-mediated production, also referred to in the literature as non-resonant, consists of an FCNC interaction producing a top quark and a new vector particle $V$ decaying into a pair of invisible DM particles, as shown in the diagrams in Figures~\ref{img:feyn_nonres_1} and \ref{img:feyn_nonres_2}. This process is described by the following Lagrangian~\cite{Andrea_2011,Abercrombie_2020,PhysRevD.86.034008,FeynRulesMonotops}:

\begin{equation}
\Lagr = aV_\mu \bar{u}\gamma^\mu P_\text{R}t + g_\chi V_\mu \bar{\chi}\gamma^\mu \chi + \textrm{h.c.},
\end{equation}

where the massive vector boson $V$ is coupled to the DM particles $\chi$. The strength of this coupling is controlled by the parameter $g_\chi$, while the $P_\text{R}$ operator is the right-handed chirality projector. The parameter $a$ stands for the coupling constant of the vector boson $V$ to the top-quark $t$ and up-type quarks $u$, and $\gamma^\mu$ are the Dirac matrices.

The scalar-mediated case, also referred to as resonant, corresponds to the production of a coloured $2/3e$-charged scalar $\phi$ that decays into a top quark and a spin-$1/2$ DM particle $\chi$~\cite{PhysRevD.86.034008}. A representative Feynman diagram of this process is shown in Figure~\ref{img:feyn_res}. The Lagrangian that describes this process is the following~\cite{Andrea_2011,Abercrombie_2020,PhysRevD.86.034008,FeynRulesMonotops}:

\begin{equation}
\Lagr = \lambda_q\phi\bar{d}^c P_R s + y_\chi\phi\bar{t}P_R\chi + \textrm{  h.c.},
\end{equation}

where $\lambda_q$ is the coupling strength of the $\phi$ scalar with $d$- and $s$-quarks, $y_\chi$ is the coupling strength of the scalar $\phi$ with the DM particle $\chi$ and the top quark, and the superscript $c$ stands for the charge conjugate.

\subsection{Production of a single vector-like $T$ quark}
\label{subsec:vlq}
\begin{figure}[h]
\centering
\subfloat[]{\includegraphics[width=0.35\textwidth]{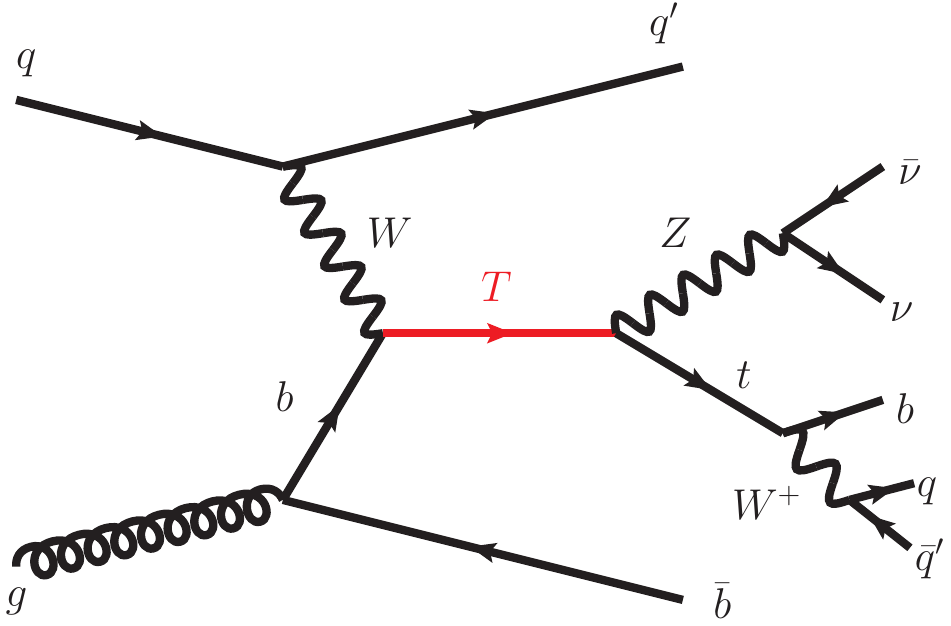}} $\quad\quad$
\subfloat[]{\includegraphics[width=0.3\textwidth]{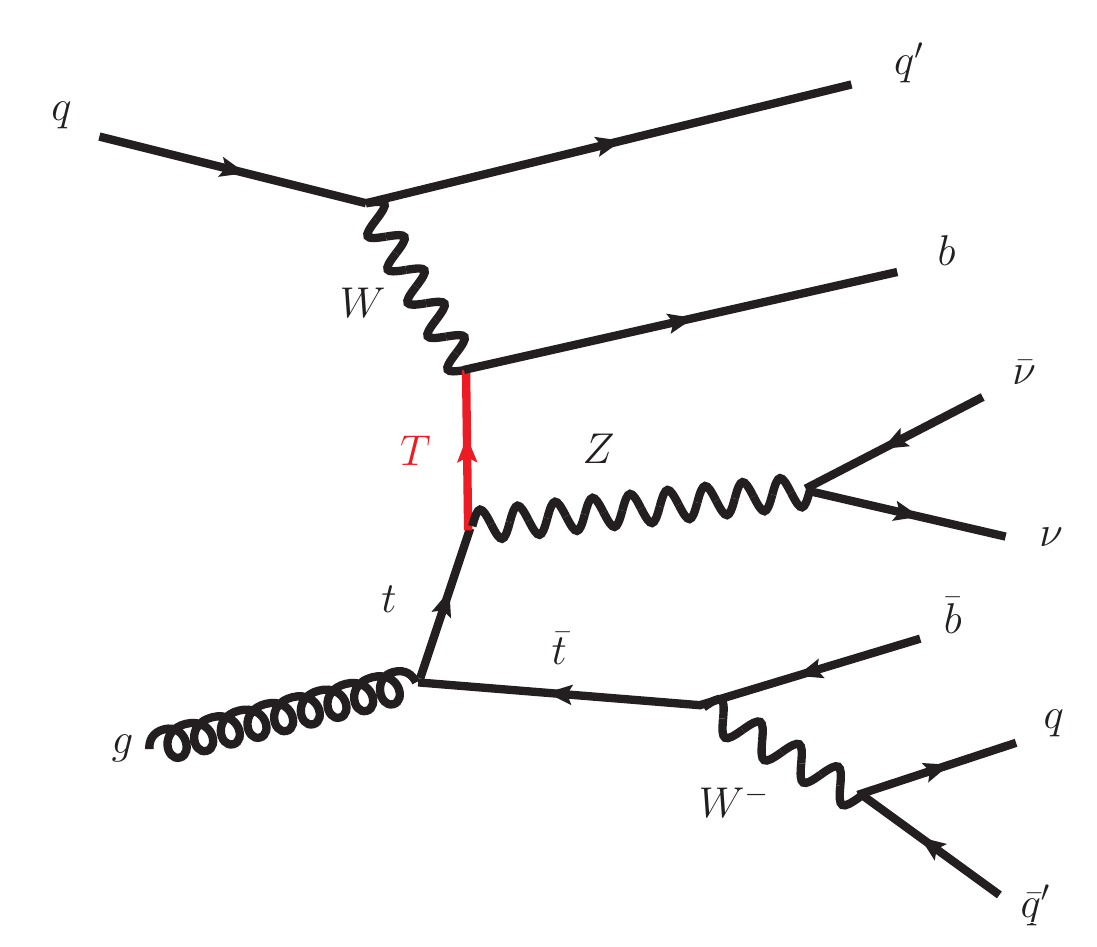}}
\caption{Representative Feynman diagrams for the (a) resonant and (b) non-resonant production of a single vector-like $T$ quark.}
\label{fig:feynman-diagrams-vlq}
\end{figure}

The production of a single vector-like $T$ quark can occur via $WTb$ and $ZTt$ vertices. However, the contribution via $ZTt$ vertex is highly suppressed by the requirement of a top quark in the initial state. Therefore, this paper considers only $WTb$ production of the vector-like $T$ quark. The dominant process in the $WTb$ associated mode is the resonant production. The non-resonant production is also considered. Vector-like $T$ quarks decay into $Wb$, $tZ$ and $tH$ with a branching ratio that depends on the considered model~\cite{Aguilar_Saavedra_2009,Matsedonskyi_2014,FeynRulesVLQ}. The respective diagrams are shown in Figure~\ref{fig:feynman-diagrams-vlq}.

The mass of the $T$~quark, $m_T$, and the overall coupling factor, $\kappa_T$, to the SM $W$~boson, $Z$~boson, and Higgs~boson are unknown parameters~\cite{Buchkremer_2013,Matsedonskyi_2014}. The overall coupling factor controls both the production cross-section and the resonance width of the $T$~quark, $\Gamma_T$. For a vector-like $T$ quark with mass \mT, the relative width (\GM) of the VLQ resonance scales quadratically with both \kappat and \mT~\cite{Buchkremer_2013}. There are also three additional parameters, $\xi_W$, $\xi_Z$ and $\xi_H$, which determine the $T$-quark branching ratios to each of the SM bosons -- $\mathcal{B}(T\to Wb)$, $\mathcal{B}(T\to tZ)$ and $\mathcal{B}(T\to tH)$, respectively. The relation $\xi_W + \xi_Z + \xi_H = 1$ is assumed, with $\xi_Z = \xi_H$. Consequently, any value of $\xi_W$ fully determines the value of $\xi_Z$ and the branching ratio of the $T\to tZ$ decay, targeted by the analysis. Since the $\xi_W<0.5$ region is dominated by the suppressed $ZTt$ production vertex, only the $\xi_W>0.5$ space is explored.

In case of a $T\to tZ$ decay, with a subsequent decay of the $Z$ boson into a pair of neutrinos, the event would have the \emph{mono-top} signature. Differently from the production of the top quark in association with DM particles, the single vector-like $T$ quark production expects additional quarks in the final state, featuring at least one jet emitted at a small angle to the beam line.

%

%
\section{ATLAS detector}
\label{sec:atlas}

%
\newcommand{\AtlasCoordFootnote}{%
ATLAS uses a right-handed coordinate system with its origin at the nominal interaction point (IP)
in the centre of the detector and the \(z\)-axis along the beam pipe.
The \(x\)-axis points from the IP to the centre of the LHC ring,
and the \(y\)-axis points upwards.
Cylindrical coordinates \((r,\phi)\) are used in the transverse plane,
\(\phi\) being the azimuthal angle around the \(z\)-axis. The pseudorapidity is defined in terms of the polar angle \(\theta\) as \(\eta = -\ln \tan(\theta/2)\).  Rapidity is defined as $y = 0.5 \ln\left[ (E + p_z ) / E - p_z)\right]$, where $E$ denotes the energy and $p_z$ is the component of the momentum along the beam direction.  Angular distance is measured in units of \(\Delta R \equiv \sqrt{(\Delta\eta)^{2} + (\Delta\phi)^{2}}\).}

The ATLAS detector~\cite{PERF-2007-01} at the LHC covers nearly the entire solid angle around the collision point.\footnote{\AtlasCoordFootnote}
It consists of an inner tracking detector surrounded by a thin superconducting solenoid, electromagnetic and hadronic calorimeters,
and a muon spectrometer incorporating three large superconducting air-core toroidal magnets.

The inner-detector system (ID) is immersed in a \qty{2}{\tesla} axial magnetic field
and provides charged-particle tracking in the range \(|\eta| < 2.5\).
The high-granularity silicon pixel detector covers the vertex region and typically provides four measurements per track,
the first hit generally being in the insertable B-layer (IBL) installed before Run~2~\cite{ATLAS-TDR-19,PIX-2018-001}.
It is followed by the SemiConductor Tracker (SCT), which usually provides eight measurements per track.
These silicon detectors are complemented by the transition radiation tracker (TRT),
which enables radially extended track reconstruction up to \(|\eta| = 2.0\).
The TRT also provides electron identification information
based on the fraction of hits (typically 30 in total) above a higher energy-deposit threshold corresponding to transition radiation.

The calorimeter system covers the pseudorapidity range \(|\eta| < 4.9\).
Within the region \(|\eta|< 3.2\), electromagnetic calorimetry is provided by barrel and
endcap high-granularity lead/liquid-argon (LAr) calorimeters,
with an additional thin LAr presampler covering \(|\eta| < 1.8\)
to correct for energy loss in material upstream of the calorimeters.
Hadronic calorimetry is provided by the steel/scintillator-tile calorimeter,
segmented into three barrel structures within \(|\eta| < 1.7\), and two copper/LAr hadronic endcap calorimeters.
The solid angle coverage is completed with forward copper/LAr and tungsten/LAr calorimeter modules
optimised for electromagnetic and hadronic energy measurements respectively.

The muon spectrometer (MS) comprises separate trigger and
high-precision tracking chambers measuring the deflection of muons in a magnetic field generated by the superconducting air-core toroidal magnets.
The field integral of the toroids ranges between \num{2.0} and \qty{6.0}{\tesla\metre}
across most of the detector.
Three layers of precision chambers, each consisting of layers of monitored drift tubes, cover the region \(|\eta| < 2.7\),
complemented by cathode-strip chambers in the forward region, where the background is highest.
The muon trigger system covers the range \(|\eta| < 2.4\) with resistive-plate chambers in the barrel, and thin-gap chambers in the endcap regions.

The luminosity is measured mainly by the LUCID--2~\cite{LUCID2} detector that records Cherenkov light produced in the quartz windows of photomultipliers located close to the beampipe.

Events are selected by the first-level trigger system implemented in custom hardware,
followed by selections made by algorithms implemented in software in the high-level trigger~\cite{TRIG-2016-01}.
The first-level trigger accepts events from the \qty{40}{\MHz} bunch crossings at a rate below \qty{100}{\kHz},
which the high-level trigger further reduces in order to record complete events to disk at about \qty{1}{\kHz}.

A software suite~\cite{ATL-SOFT-PUB-2021-001} is used in data simulation, in the reconstruction
and analysis of real and simulated data, in detector operations, and in the trigger and data acquisition
systems of the experiment.


%

%
\section{Data and simulated samples}
\label{sec:samples}

This analysis uses data from $pp$ collisions at a centre-of-mass energy of $\sqrt{s}=13$~\TeV collected by the ATLAS detector during the LHC Run 2 from 2015 to 2018.  In this data taking period, the LHC operated with 25~ns proton bunch spacing with an average number of collisions per bunch crossing $\langle\mu\rangle$ of $34$. Only data-taking periods in which all the necessary components for the analysis were functional are included, corresponding to an integrated luminosity of $139$~fb$^{-1}$.

Events are required to pass at least one of the triggers that select events with high missing transverse momentum,
with thresholds between 90 and 120~\GeV depending on the data taking year~\cite{TRIG-2019-01}.
All the triggers are fully efficient for events with offline $\MET>250$~\GeV, as selected for analysis.

Samples of simulated signal and background processes are produced using different Monte Carlo (MC) event generators including parton shower and hadronisation models.

Signal events for both the scalar and vector mediator DM scenarios are generated with the simplified model described in Section~\ref{sec:signal}, using \MGNLO v2.8.1 and v2.9.5~\cite{Alwall:2011uj} with \FEYNRULES~2.0~\cite{Alloul_2014}. The generation is done at leading-order (LO) accuracy in QCD using the NNPDF3.0LO parton distribution function (PDF) set~\cite{Ball:2014uwa}. The generator is interfaced with \PYTHIA{8.245}~\cite{Sjostrand:2014zea} using the A14 set of tuned parameters ({\textit{tune}})~\cite{ATL-PHYS-PUB-2014-021} and the NNPDF2.3LO PDF set~\cite{Ball:2012cx}. %

A total number of 40 (19) samples are simulated to cover the four dimensions in the parameter space -- two couplings plus two masses -- of the scalar (vector) mediator DM production. The larger number of simulations of the scalar mediator DM production is due to the non-negligible dependence of the event kinematics on $m_\phi$ and $\lambda_q$ that required the production of more samples to properly control and validate the signal predictions. The scalar DM samples were simulated in a grid of the parameter space encompassed by the following ranges: $m_\phi\in[2500,6000]$~\GeV, $m_\chi\in[500,5500]$~\GeV, $\lambda_q\in[0.2,1]$ and $y_\chi\in[0.2,0.6]$. For the vector-mediated signals, samples were simulated with the following parameter range: $m_V\in[1250,3500]$~\GeV, $m_\chi\in[400,1200]$~\GeV, $a\in[0.1,0.5]$ and $g_\chi\in[0.2,1]$. The parameters of the DM mediators and the VLQ signals are chosen to scan masses higher than the previous excluded limits~\cite{EXOT-2017-16}.

The signal predictions in points of the parameter space that are not simulated are obtained through a reweighting procedure that exploits the mild dependence of the event kinematics on the other parameters. The reweighting procedure consists of assigning weights to the available simulated samples (called \emph{reference}) to reproduce the expected kinematics and yields in a different (but close) point of the model parameter space, the \emph{target}. This method makes use of an extensive set of samples simulated at particle-level that cover the four model parameteres for both the scalar and vector mediator DM production with a higher granularity than the set of reconstruction-level samples. The mentioned weights $\boldsymbol{w}_\text{rw}$, evaluated entirely with particle-level simulations, reproduce the \emph{target} kinematics from the \emph{reference} sample. They take into account the differences in acceptance, cross-sections at LO and kinematic shapes between the \emph{reference} and \emph{target} samples, as summarised in the equation:

\begin{equation}
w^{i}_\text{rw}= \overbrace{ \frac{\epsilon_\text{target}}{\epsilon_\text{reference}}}^{\text{Acceptance}} \times \overbrace{ \frac{\sigma_\text{target}}{\sigma_\text{reference}} }^\text{Cross-section} \times \overbrace{ \frac{y^i_\text{target}}{y^i_\text{reference}}}^\text{Shapes}.
\end{equation}

The acceptance term is determined by applying the event preselection, defined in Section~\ref{sec:selection}, on particle-level simulation.~\footnote{Particle-level objects in Monte Carlo simulation are reconstructed from stable particles (with proper lifetimes $c\tau \geq 10$~mm). Particle-level jets are clustered from visible stable final-state particles, excluding prompt leptons, using the anti-$k_t$ clustering.
Particle-level top jets are reconstructed using the anti-$k_t$ algorithm with radius parameter $R=1.0$ and are required to be close to a top-quark ($\Delta R< 1$).}
It is measured as the ratio of the preselection efficiencies for the \emph{target} sample ($\epsilon_\text{target}$) and the \emph{reference} sample ($\epsilon_\text{reference}$).
The change in the cross-section between the \emph{target} ($\sigma_\text{target}$) and the \emph{reference}  ($\sigma_\text{reference}$) is accounted for by the ratio of the two values. The change in event kinematics is estimated with the ratio of the yields of the normalised binned \met distribution for the \emph{reference} ($y^i_\text{reference}$) and \emph{target} ($y^i_\text{target}$) samples, where $i$ is the bin corresponding to the truth \met (see Section~\ref{sec:reco}) of the \emph{reference} and \emph{target} events. The reweighting based on the \met observable has shown to be enough to accurately reproduce the distributions of the other kinematic variables relevant to the analysis.

The single production of a singlet vector-like $T$ quark is simulated at LO in QCD using the corresponding model~\cite{Buchkremer_2013} implemented in \FEYNRULES[2.0] with \MGNLO[v.2.8.1] using the NNPDF3.0LO PDF set. The simulation comprises all the tree-level processes,  including the non-resonant production. The generator is interfaced with \PYTHIA[8.244] using the NNPDF2.3LO PDF set with the A14 tune. The samples are generated for masses of the vector-like $T$ quark ranging from $1.1$ to $2.7$~\TeV in steps of 0.2~\TeV, assuming $\kappa_T=0.5$. Internal weights provided by \MGNLO are used to obtain predictions
at masses $100$~\GeV smaller than the generated ones
and at different values of the coupling, ranging from $0.1$ to $1.6$ in steps of $0.05$ for $\kappa_T<0.5$ and $0.1$ for larger $\kappa_T$.
The mass of the vector-like quark, $m_T$, strongly affects the boost of the top quark, while small, but non-negligible, changes of the event kinematics are observed for different values of the $\kappa_T$ coupling.
Signal samples are normalised to next-to-leading-order (NLO) cross-sections in QCD  that assume the narrow-width approximation for the $T$-quark decay~\cite{CACCIAPAGLIA2019206}. Finite-width effects on the cross-sections are accounted for by applying a correction factor~\cite{PhysRevD.101.115027}, ranging from a few per cent up to approximately $30\%$ at large values of the width.
The production of a top-antitop-quark pair (\ttbar) is simulated at NLO using \POWHEGBOX[v2]~\cite{Nason:2004rx,Frixione:2007vw,Alioli:2010xd}, with the NNPDF3.0NNLO PDF set, interfaced with \PYTHIA[8.210] using NNPDF2.3LO PDF set and the A14 set of tuned parameters.
The events are normalised to the next-to-next-to-leading-order (NNLO) QCD cross-section, including resummation of soft gluon emissions at next-to-next-to-leading logarithmic (NNLL) accuracy calculated using the \textsc{Top++2.0}~\cite{Czakon:2011xx} software.

The production of a vector boson ($V=W,Z$) in association with jets (\Vjets) is simulated with the \SHERPA[v2.2.1]~\cite{Gleisberg:2008ta} generator in which the matrix element is accurate at NLO up to two emitted partons while at LO up to four additional partons. These calculations are performed with the Comix~\cite{Gleisberg:2008fv} and OpenLoops~\cite{Cascioli:2011va} libraries. The matching of the parton-shower and hadronisation algorithm with the matrix-element generator~\cite{Hoeche:2011fd} is employed for different jet multiplicities, which are then merged into an inclusive sample using an improved CKKW matching procedure~\cite{Catani:2001cc,Hoeche:2009rj} that is extended to NLO accuracy using the MEPS@NLO prescription~\cite{Hoeche:2012yf} via a set of parameters provided by the authors.
The NNPDF3.0NNLO PDF set is used~\cite{Ball:2014uwa}. Both processes are normalised to the NNLO cross-section~\cite{Anastasiou:2003ds}.

Single-top $tW$ production is simulated using the \POWHEGBOX[v2] generator at NLO in QCD using the five-flavour scheme and the NNPDF3.0NNLO PDF set. The diagram removal scheme~\cite{Frixione:2008yi} is used to remove the overlap with \ttbar production. Single-top $t$-channel production is modelled with \POWHEGBOX[v2] generator at NLO in QCD using the four-flavour scheme. The NNPDF3.0NLOnf4 PDF set is used. Single-top $s$-channel  production is simulated using the \POWHEGBOX[v2] generator at NLO in QCD using the five-flavour scheme. The NNPDF3.0NNLO PDF set is used. In all the single-top-quark processes, the generator is interfaced with \PYTHIA[8.230] using the A14 tune and the NNPDF2.3LO PDF set.
Simulated samples are normalised to the NLO cross-section in QCD including the resummation of soft-gluon emission corrections~\cite{BORDES199523,PhysRevD.58.094021,PhysRevD.56.5919,PhysRevD.83.091503,PhysRevD.81.054028,PhysRevD.82.054018}.

The production of \ttbar in association with a $W$ or $Z$ boson is simulated with \MGNLO[v2.3.3]~\cite{Alwall:2014hca} at NLO QCD accuracy using the NNPDF3.0NNLO PDF set. The generator is interfaced with \PYTHIA[8.230] using the A14 tune and the NNPDF2.3LO PDF set.
The samples are normalised to NLO QCD and electroweak (EW) cross-sections. The \ttbar{$Z$} cross-section is additionally corrected by $Z$ boson off-shell contributions.

Events from \ttbar{$H$} production are simulated at NLO QCD accuracy with \POWHEGBOX[v2] generator with the NNPDF3.0NNLO PDF set. The generator is interfaced with \PYTHIA[8.230] using the A14 tune and the NNPDF2.3LO PDF set.
The cross-section is calculated at NLO QCD and NLO EW accuracy.

Diboson events are simulated with the \SHERPA[v2.2.1] generator, using the same prescription as in the production of single bosons in association with jets.

The production of $tZq$ is simulated with \MGNLO[v2.3.3] at NLO QCD accuracy using the NNPDF3.0NNLO PDF set. The production of $tWZ$ with $Z\rightarrow\nu\bar{\nu}$ is simulated at NLO QCD accuracy with \MGNLO[v2.6.7] using the NNPDF2.3LO PDF. %

The simulation of bottom and charm hadron decays is done using \EVTGEN[]~\cite{Lange:2001uf} in all signal samples and all background samples with the exception \Vjets and dibosons (generated with \SHERPA[v2.2.1]). The generation also includes the effect of multiple $pp$ interactions per bunch crossing (pile-up), and the effect on the detector response due to interactions from bunch crossings before or after the one containing the hard interaction~\cite{ATL-PHYS-PUB-2016-017}.
All the simulated samples are processed with a detailed simulation of the ATLAS detector~\cite{SOFT-2010-01} using \textsc{Geant4}~\cite{Agostinelli:2002hh}.
%


%

%
\section{Event reconstruction and object selection}
\label{sec:reco}

All events are required to contain a primary vertex with at least two associated tracks with transverse momentum $\pT>0.5$~\GeV. The vertex with the highest $\sum \pT^2$\ of the associated tracks is taken as the hard-scatter primary vertex.

Electron candidates are identified from high-quality inner-detector tracks matched to calorimeter
energy deposits consistent with an electromagnetic shower.
The energy deposits have to form a cluster with transverse energy $\et > 25$~\GeV\ and $|\eta| < 2.47$, and
be outside the transition region $1.37 \leq |\eta| \leq 1.52$\ between the barrel and endcap calorimeters. A tight likelihood-based requirement is used to reject fake-electron candidates~\cite{EGAM-2018-01}. Electrons are not required to be isolated to maximise the rejection of background events when applying the lepton veto preselection described in the next section.

Muon candidates are reconstructed using high-quality inner-detector tracks combined with tracks
reconstructed in the muon spectrometer.
Only muon candidates satisfying `medium' identification criteria~\cite{MUON-2018-03}, with $\pt > 25$~\GeV\
and $|\eta| < 2.5$, are considered. No isolation criteria are applied to improve the background rejection with the lepton veto (see Section~\ref{sec:selection}).

Jet candidates are reconstructed using the \antikt~\cite{Cacciari:2008gp} jet algorithm with radius parameter $R=0.4$ ($1.0$) for small-$R$ (large-$R$) jets, as implemented in the FastJet~\cite{Fastjet} package.
The large-$R$ jets are formed from topological clusters in the calorimeter calibrated using the local calibration method described in Ref.~\cite{PERF-2014-07}, while the small-$R$ jets are reconstructed from particle-flow objects~\cite{PERF-2015-09}. Both are calibrated by applying a jet energy scale derived from 13~\TeV\ data and simulation~\cite{JETM-2018-02, JETM-2018-05}.

After calibration, small-$R$ jets are considered if they fulfil $p_{T} > 25$~\GeV and $|\eta| < 4.5$. Jets in the region $2.5 < |\eta| < 4.5$ are considered forward jets. To suppress jets originating from pile-up collisions, a requirement on the jet-vertex tagger (JVT)~\cite{PERF-2014-03} discriminant is applied for jets with \pt below 60~\GeV and $|\eta| < 2.4$.
Jets that are close to an electron with a distance $\Delta R_{y, \phi} \equiv \sqrt{\left(\Delta y\right)^2 + \left(\Delta \phi\right)^2} < 0.2$ are removed to avoid double counting the energy of the electron. Jets with less than three tracks are removed if their distance to a muon is $\Delta R_{y, \phi}  < 0.2$ or one of their tracks is part of the muon. %
Electrons or muons with a distance $\Delta R_{y, \phi} < 0.4$ from any of the remaining jets are removed.

Small-$R$ jets that contain $b$-hadrons ($b$-jets) are identified by the use of the DL1r multivariate algorithm~\cite{FTAG-2019-07}. These jets are hereafter referred to as $b$-tagged jets. The selected working point results in an efficiency of $77\%$ per $b$-jet and rejection factors of 6 and 200 for $c$- and light-flavour jets, respectively, as measured in simulated $\ttbar$ events.

A trimming algorithm~\cite{trimming} with parameters $R_{\mathrm{sub}} = 0.2$ and $f_{\mathrm{cut}} = 0.05$\ is applied to
the large-$R$ jets to mitigate the impact of gluon radiation and \pileup\ effects. Trimmed large-$R$ jets are considered if they fulfil $p_{T} >250$~\GeV\ and $|\eta| < 2.0$.
To identify a large-$R$ jet originated from hadronically decaying top quarks produced with a large Lorentz boost ({\textit top-tagged}), the \emph{contained} (similar to \emph{medium}) top-quark tagging algorithm is used~\cite{JETM-2018-03,ATL-PHYS-PUB-2020-017}.
The top-quark tagging relies on a deep neural network (DNN) that uses jet kinematics and substructure variables
(e.g., energy flow) as inputs~\cite{JETM-2018-03,ATL-PHYS-PUB-2020-017}.
The \pT-dependent requirements on the DNN score provide 50\%\ top-quark-tagging
efficiency, as measured in inclusive \ttbar\ events where the
top-quark decay products are considered to be contained within the large-$R$ jet, and a corresponding background rejection up to 70 (95) for multijet ($\gamma$+jets) events depending on the jet $\pt$~\cite{JETM-2018-03,ATL-PHYS-PUB-2020-017}.

The missing transverse momentum vector $p_{\text{T}}^{\text{miss}}\xspace$ is reconstructed from the negative vector sum of momenta of calibrated leptons, small-$R$ jets and a soft term built from all tracks that are associated with the primary vertex but not with these objects~\cite{JETM-2020-03}. Its magnitude is denoted by $\MET$. The truth $\MET$ used in the reweighting procedure described in Section~\ref{sec:samples} is calculated from the truth $p_{\text{T}}$ of invisible particles.


%

%
\FloatBarrier
\section{Event selection and background estimate}
\label{sec:selection}

The experimental signature expected in the considered DM and vector-like $T$-quark models is the presence of a top quark and significant missing transverse momentum.
As the top quark is expected to be produced with a large Lorentz boost, its hadronic decay products can be collimated and reconstructed into a large-$R$ jet. Preselected events are required to contain zero leptons and at least one large-$R$ jet. The leading large-$R$ jet ($J$) is then required to be top-tagged and have a $\pT$ between $350$~\GeV and $2500$~\GeV, and a mass between $40$~\GeV and $600$~\GeV, corresponding to the region where the top-tagging algorithm is calibrated.
To reduce the number of multijet background events to a negligible level, the $\met$ is required to be larger than $250$~\GeV. Quality criteria are imposed to reject events that contain any jets arising from non-collision sources or detector noise~\cite{ATLAS-CONF-2015-029}. To remove the contribution from mis-reconstructed $\met$ aligned to badly reconstructed jets,  the minimum distance in the transverse plane between any reconstructed small-$R$ jet ($j$) and the missing transverse momentum, $\Delta\phi_\text{min} (j,\met)$, is required to be larger than 0.2.

\begin{figure}[t]\centering
\includegraphics[width=0.8\textwidth]{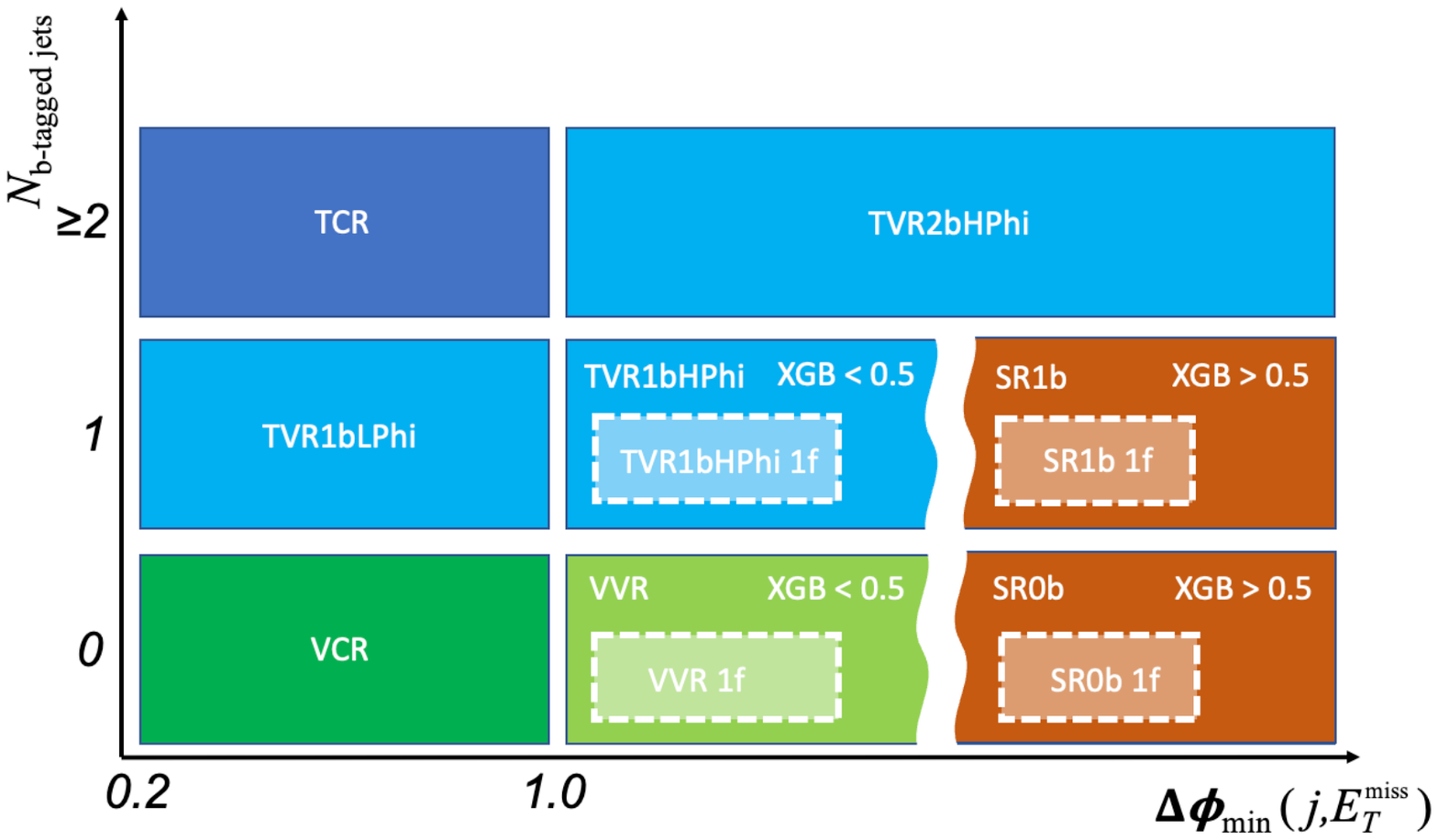}
\caption{Schematic representation of the control, validation and signal regions. Regions are defined in terms of $b$-tagged jet multiplicity $N_{b\text{-tagged jets}}$ and the minimum distance in the azimuthal angle between a jet and \met \dphimin. The notation ``XGB<0.5'' and ``XGB>0.5'' indicates the requirement on the \xgb~score of the validation and signal regions, respectively. These regions are schematically separated by the curly vertical line. The ``1f''  label stands for requiring at least one forward jet in the event. The selections used to define the regions are described in the text and in Table~\ref{tab:region_summary}. }
\label{fig:monoregions}
\end{figure}

Additional requirements are used to define several orthogonal regions: signal regions (SRs) optimised to maximise the sensitivity to the target signal models; control regions (CRs) designed to measure the normalisation of the main contributing background processes; and validation regions (VRs) to check the background modelling in regions kinematically closer to the signal regions. The control and validation regions are defined to have a negligible signal contribution. The requirements that define the analysis regions are listed in Table~\ref{tab:region_summary}, sketched in Figure~\ref{fig:monoregions} and described as follows.

\begin{table}[t]
\caption{Summary of the event selections used to define the signal, control and validation regions. The signal regions are denoted by SR0b and SR1b, the \ttbar (\vjets) dominated control regions are denoted by TCR (VCR) and the validation regions enhanced in \ttbar (\vjets) are labelled as TVR1bLPhi, TVR1bHPhi and TVR2bHPhi (VVR). The ``(1f)'' notation in the signal and validation region rows indicates the additional requirement of at least one forward jet in the event, which is applied for the search of a single vector-like $T$ quark. All regions are required in addition to contain zero leptons in the final state, $\met\geq250$~\GeV and at least one top-tagged large-$R$ jets with $\pT \in [350,2500]$~\GeV and a mass $\in [40, 600]$~\GeV. The number of $b$-tagged (forward) jets required is indicated by $N_{b\text{-tagged jets}}$ ($N_\text{forward-jets}$). The symbol \emph{-} indicates that no requirement on the variable is applied.}
\centering
\begin{tabular}{@{}lcccc@{}}

\toprule
&  $N_{b\text{-tagged jets}}$  & \dphimin & \xgb score & $N_\text{forward jets}$ \\
\midrule
TCR        & $\geq 2$ &   $\in [0.2, 1]$   & --    & --     \\
TVR1bLPhi     & $1$ &   $\in [0.2, 1]$   & --    & --  \\
TVR1bHPhi (1f)     & $1$ &   $\geq 1$   & $<0.5$    & --\,\,\,($\geq 1$)  \\
TVR2bHPhi   & $\geq 2$ &   $\geq 1$   & --    & --    \\  [2mm]
VCR      & $0$ &   $\in [0.2, 1]$   & --    & --     \\
VVR (1f)  & $0$ &    $\geq 1$   & $<0.5$     & --\,\,\,($\geq 1$)    \\
\hline
SR0b (1f) & $0$ &    $\geq 1$   & $\geq 0.5$     & --\,\,\,($\geq 1$)    \\
SR1b (1f) & $1$ &    $\geq 1$   & $\geq 0.5$     & --\,\,\,($\geq 1$)   \\
\bottomrule
\end{tabular}
\label{tab:region_summary}
\end{table}

\subsection{Signal region definition}

Different SRs are defined for the three signal scenarios considered to maximise the analysis sensitivity to each model. Each signal region is defined by the output score of an \xgb classifier~\cite{10.1145/2939672.2939785} specifically trained to improve the sensitivity to each signal.

The \xgb algorithm~\cite{10.1145/2939672.2939785} combines several input observables into an extreme gradient BDT. Gradient boosting improves over a single decision tree classifier by combining a number of classifiers iteratively enhanced. Three \xgb classifiers are trained separately with preselected simulated background and signal events for each considered model - scalar and vector DM, and VLQ. In the training of the classifier targeting the separation of events from scalar (vector) mediator DM production, the simulated sample with a mediator mass of $m_\phi$ = 4~\TeV ($m_V$ = 1.75~\TeV) is used, corresponding to an intermediate value of the explored masses.
The classifier targeting the identification of single vector-like $T$ production is trained using a set of samples with $m_T$ ranging from $1.5$ to $1.9$~\TeV.
This choice provides stable performances for larger masses while for lower masses, which are mostly excluded already, some degradation is observed.

The hyperparameters of each \xgb model are optimised to maximise the integral under the receiver operating characteristic curve. Similarly, the set of variables used in the training of the classifier, listed in Table~\ref{tab:xgbvars}, are selected based on their discrimination power, and good data and simulation agreement. The training of the \xgb classifier targeting the DM scalar production required fewer input variables, as signal events are mostly characterised by large \met values, of the order of the mediator mass. For the remaining signal models, the kinematic properties of signal events are more similar to those from background processes, and therefore more input features are required for the event classification. The \met is the most discriminating variable for all considered models, as it is expected to be significantly larger than for background events. The reconstructed transverse mass of the system composed of the \met and the closest $b$-tagged jet plays an important role in the discrimination against \ttbar events with a top quark decaying leptonically, for which the distribution is expected to peak at around the top-quark mass.
Variables based on angular distances of the reconstructed objects and energy balance between \met and the jets (e.g., $\Omega$) also help in the event classification.

\begin{table}[t]
\caption{List of variables used in the training of the \xgb classifier for the scalar and vector DM mediator, and VLQ models. The symbol $\checkmark$ indicates that the corresponding variable is included in the training.}
\centering
\resizebox{\textwidth}{!}{
\begin{tabular}{@{}llcccc@{}}
\toprule
Variable & Description & $\quad$ & Scalar & Vector & VLQ \\

\multicolumn{2}{c}{} & $\quad$ & DM mediator & DM mediator &  \\

\midrule
{$E_{\text{T}}^{\text{miss}}$} & \makecell[l]{\scriptsize Missing transverse momentum} & $\quad$ & $\checkmark$ & $\checkmark$ & $\checkmark$ \\[3mm]
{$\Omega$} &{\scriptsize \met and large-$R$ jet \pT balance: $ \frac{ \met - \pT(J) }{\met + \pT(J) }$}& $\quad$ & $\checkmark$ & $\checkmark$ & $\checkmark$\\[3mm]
{N$_{\text{jets}}$}  & {\scriptsize Small-$R$ jet multiplicity }& $\quad$ & $\checkmark$ & $\checkmark$ & $\checkmark$\\[3mm]
{$\Delta R_{\text{max}}$}  & \makecell[l]{  \scriptsize Maximum $\Delta{R}$ between \scriptsize  two small-$R$ jets} & $\quad$ & $\checkmark$ & $\checkmark$ & $\checkmark$\\[3mm]
{$m_{\text{T,min}}(E_{\text{T}}^{\text{miss}}$, $b$\text{-tagged jet})} & \makecell[l]{\scriptsize Transverse mass of \met and  the closest  $b$-tagged jet} & $\quad$ & $\checkmark$ & $\checkmark$ & $\checkmark$\\[3mm]
{$m_{\text{top-tagged jet}}$} & \makecell[l]{\scriptsize Mass of the large-$R$ top-tagged jet} & $\quad$ & $\checkmark$ & &  $\checkmark$ \\[3mm]
{$\Delta p_{\text{T}}$ \text{($J$, jets)}} & \makecell[l]{\scriptsize Scalar difference of large-$R$ jet \pT and  the sum \\   \scriptsize of \pT of all small-$R$ jets.} & $\quad$ & $\checkmark$ & $\checkmark$ &  \\[3mm]
$H_{\text{T}}$ & \makecell[l]{\scriptsize Sum of all small-$R$ jet \pT} & $\quad$ & & $\checkmark$ & $\checkmark$\\[3mm]
$H_{\text{T}}/E_{\text{T}}^{\text{miss}}$  & \makecell[l]{\scriptsize Ratio of  $H_{\text{T}}$ and \met} & $\quad$ & & $\checkmark$ & $\checkmark$\\[3mm]
$\Delta E(E_{\text{T}}^{\text{miss}}$, $J$) & \makecell[l]{\scriptsize Energy difference between \met and the large-$R$ jet}& $\quad$ & & $\checkmark$ & $\checkmark$\\[3mm]
$\Delta\phi(E_{\text{T}}^{\text{miss}}$, $J$) & \makecell[l]{\scriptsize Angular distance in the transverse plane  between \\ \scriptsize \met and large-$R$ jet }  & $\quad$ & & $\checkmark$ & $\checkmark$ \\[3mm]
$p_{\text{T}}($J$)$ & \makecell[l]{\scriptsize Large-$R$ jet \pT }& $\quad$ & & & $\checkmark$\\[3mm]
{$m_{\text{T}}(E_{\text{T}}^{\text{miss}}$, $J$)} & \makecell[l]{\scriptsize Transverse mass of the \met and large-$R$ jet}& $\quad$ &  &  & $\checkmark$\\[3mm]
$\Delta\phi$($b$\text{-tagged jet}, $J$) & \makecell[l]{\scriptsize Angular distance in the transverse plane between the \\ \scriptsize large-$R$ jet and the leading $b$-tagged jet} & $\quad$ &  &  & $\checkmark$\\[3mm]

\bottomrule
\end{tabular}
}
\label{tab:xgbvars}
\end{table}

For signal events, the top quark is expected to recoil against the missing transverse momentum, thus the requirement of \dphimin $>1$ is applied. This requirement rejects \ttbar events with a boosted top quark decaying hadronically and recoiling against a \bjet from the leptonic decay of the other top quark.
The signal regions for a given model are defined by requiring the corresponding $\xgb$ score to be larger than 0.5.
The resulting events are separated into two orthogonal signal regions (SR0b,SR1b) by applying criteria based on $b$-tagged jet multiplicity (see Figure~\ref{fig:monoregions}). The SR1b (SR0b) signal region must contain exactly one (zero) $b$-tagged jet. Signal events are expected to have one $b$-jet from the top-quark decay in the final state, but the SR0b is expected to account for $b$-tagging inefficiencies. The signal efficiencies of the SRs targeting each DM model depend on the specific signal parameters and are within 14\%--27\% and 3\%--9\% for the scalar and vector DM, respectively.

The signal regions targeting the search for the vector-like $T$ quark production require additionally at least one forward jet since final states with jets at a small angle relative to the beam pipe are expected. The VLQ signal selection efficiency in these SRs is 1\%--2\% depending on the model parameters.

In summary, the analysis employs six distinct regions to target the three signal models with two SRs each: SR0b and SR1b region for the DM scalar search, SR0b and SR1b region for the DM vector, and SR0b 1f and SR1b 1f targeting single vector-like $T$-quarks.

\subsection{Background estimate}
Dedicated control regions enriched in the dominant background processes are defined to constrain their normalisation to data in a profile-likelihood fit described in Section~\ref{sec:results}, while the shape distributions are taken from simulation. The rest of the background processes are estimated completely from the simulation.

The main contributing background sources to the signal region are \ttbar and \vjets production. The former represents approximately $65\%$ of the background events in SR1b,  consistent across all the defined signal regions with one $b$-tagged jet requirement. The other main sources of background in this region are \vjets ($20\%$) and single top ($10\%$), consisting mostly of $tW$ production. The SR0b is dominated by \vjets production (up to $85\%$) across all the inspected models, with minor contributions from diboson production ($10\%$) and \ttbar ($<8\%$).

Control regions enhanced in the dominant \ttbar and \vjets backgrounds are then defined by requiring $0.2<$~\dphimin~$<1$ to reduce the signal contribution below $5\%$ for all the considered models.
The \ttbar control region (referred to as TCR) is defined by requiring, in addition, at least two $b$-tagged jets.
This region is enriched in \ttbar production with at least one top quark decaying into leptons that are not reconstructed. This provides a significant source of \met recoiling against the top quark, similar to what is found in the SR1b region, where this source of background is dominant.
The control region enriched in \vjets~(referred to as VCR) requires zero $b$-tagged jets in the final state. In this region, the fractions of $W$+jets processes in which the $W$ boson decays into $\tau$-leptons, and of $Z$+jets processes with the $Z$ boson decaying into neutrinos, both leading to a hadronic final state with significant \met, are similar to those expected in the SR0b signal region.

Validation regions are defined to check the modelling of the main background processes in regions kinematically closer to the signal regions, maintaining low signal contamination ($<10\%$). The \ttbar modelling is checked in three validation regions, approaching the SR1b from two different paths in the plane shown in Figure~\ref{fig:monoregions}. The TVR2bHPhi region extends the TCR (with at least two $b$-tagged jets) to higher values of \dphimin, by requiring \dphimin~$>1$. The TVR1bLPhi region is defined with events fulfilling  $0.2<$~\dphimin~$<1$ but with a single $b$-tagged jet in the final state to validate the \ttbar modelling with a reduced fraction of additional jets originated from heavy flavour quarks, as in SR1b. Events in these two regions are not required to fulfil any requirement in the \xgb score. Finally, the TVR1bHPhi region extends the TVR1bLPhi to \dphimin~$>1$, but \xgb score $<0.5$ is required to not overlap with SR1b.
The \vjets modelling is tested in the VVR region, which requires no $b$-tagged jets in the final state, \dphimin~~$>1$ and \xgb score $<0.5$ to ensure orthogonality with the SR0b.
Two specific validation regions for the search targeting the vector-like $T$ quark production are defined by adding the requirement of at least one forward jet in the final state to the selections that define the VVR and TVR1bHPhi, called VVR1f and TVR1bHPhi1f, respectively.

%

%

%
%
%
%
%
%
%


%

%
\FloatBarrier
\section{Systematic uncertainties}
\label{sec:systematics}

Systematic uncertainties of theoretical and instrumental sources affect the predictions of background and signal in the analysis regions. The effect of these uncertainties is included in the likelihood fit as additional nuisance parameters (NP) that are measured simultaneously with the normalisations of the main background processes and the signal strength.

Several sources of instrumental uncertainties are considered. These include systematic uncertainties in the online event selection efficiency using the high-\met trigger, and also in the reconstruction and calibration of the objects analysed. Moreover, the normalisation of the simulated samples to the integrated luminosity collected in Run~2 has an uncertainty of $1.7\%$~\cite{ATLAS-CONF-2019-021}, obtained using the LUCID-2 detector~\cite{LUCID2} for the primary luminosity measurements. Among all the detector-related uncertainties, the most significant arise from the measurements of jet properties and tagging efficiencies.

Uncertainties associated with jets are due to the jet energy scale (JES) and jet energy resolution (JER)~\cite{JETM-2018-02, JETM-2018-05}. For large-$R$ jets, uncertainties arising from the jet mass scale (JMS), and jet mass resolution (JMR) are also considered. They are derived from observations of the $W$ boson and top-quark masses in semileptonic \ttbar\
events and by a double-ratio method that compares the calorimeter-to-tracker response ratio between data and simulation~\cite{JETM-2018-02}.
For small-$R$ jets, additional uncertainties related to the JVT requirement are also evaluated, arising from the correction factors used to match the efficiencies in the MC samples to data~\cite{PERF-2014-03}.

The $b$-quark tagging efficiencies in simulation, and the charm and light jet mistag rates, are corrected to match the efficiencies in data~\cite{FTAG-2018-01, FTAG-2020-08, FTAG-2019-02}. The efficiency for tagging $b$-jets is measured in data using \ttbar events~\cite{FTAG-2018-01}. Uncertainties arising in the evaluation of the efficiencies are propagated to the correction factors.

The efficiency and rejection power of the DNN top-quark tagger is measured in data and correction
factors are applied to MC events to match the measured efficiencies~\cite{ATL-PHYS-PUB-2020-017}.
These corrections take into account the correlations between the tagging efficiencies and
other jet observables such as the jet energy and mass. The uncertainties in these corrections are treated as systematic uncertainties.

A set of uncertainties that parametrise the lack of knowledge on the background prediction of theoretical nature are considered. These uncertainties are evaluated either from variations of parameter settings in the event generation or from a direct comparison to alternative simulated samples.

Uncertainties in the modelling of the \ttbar background come from the choice of NLO-matching
method, the parton shower and hadronisation modelling, the amount of
additional gluon radiation, missing higher orders in the generation and the choice of PDF. The NLO-matching
uncertainty is estimated by a direct comparison of the \ttbar simulation with an alternative sample using \MGNLO. Similarly, the uncertainty in the parton shower and hadronisation model is estimated from an alternative sample using \HERWIG{7.1}~\cite{Bellm:2017jjp}.
The uncertainty associated with the effects of missing QCD higher orders is estimated by independently varying the renormalisation and factorisation scales by factors of $2.0$ and $0.5$. The impact of initial-state radiation (ISR) is estimated by varying $\alpha_s$ in the A14 tune.  Similarly, the uncertainty related to final-state radiation (FSR) is assessed by varying the renormalisation scale for the final-state radiation by a factor of $2.0$ and $0.5$ in the parton shower algorithm.
The uncertainties in the modelling of additional gluon radiation and missing higher orders are also estimated for all single-top processes using the same procedure as in the \ttbar background. The corresponding NPs are treated as uncorrelated among  single-top and \ttbar processes.  The choice of the scheme to account for the interference between $tW$ and \ttbar is another source of uncertainty in the background prediction and is estimated by comparing predictions using diagram removal and diagram subtraction schemes ~\cite{Frixione:2008yi}.

Uncertainties in the modelling of weak boson production in association with jets, \wjets and \zjets, arise from missing higher orders in perturbation theory, resummation of the gluon radiation, matching of the  parton shower, modelling of heavy-flavour quark production and the choice of PDF.  The effect of the modelling uncertainties in \wjets and \zjets are treated as uncorrelated NPs in the profile likelihood fit. The uncertainty due to missing QCD higher orders is estimated by independently varying the renormalisation and factorisation scales by a factor of two and taking the largest observed deviation from the nominal prediction.
A similar procedure is used to assess the uncertainties associated to the resummation of gluon radiation and matching with the parton shower, as they are estimated by varying by a factor of two the corresponding scales.
Two normalisation uncertainties of $30\%$ are assigned to \vjets events with \bjets and $c$-jets, respectively,  to cover possible discrepancies from simulations in case of heavy-flavour production~\cite{ATL-PHYS-PUB-2017-006}. The presence of $b$-jet ($c$-jets) in the final state is determined by the presence of $B$-hadrons ($D$-hadrons) in the truth record of the simulated events, which are associated with reconstructed jets.  The same procedure is used to estimate the modelling uncertainties in the generation of diboson production (except for the resummation and parton shower matching uncertainties). %

Uncertainties due to the PDF choice in the simulations on \ttbar, \vjets, single-top and diboson production simulations are estimated by following the PDF4LHC~\cite{Butterworth:2015oua} recommendation.

Uncertainties in the cross-sections of the background processes whose normalisations are not measured in the control regions are also included. These background processes are single-top~\cite{Kant:2014oha, Kidonakis:2010ux,Kidonakis:2013zqa}, diboson~\cite{ATL-PHYS-PUB-2016-002} and \ttbar{$+X$}~\cite{deFlorian:2016spz,ATL-PHYS-PUB-2016-005}. The respective uncertainties are 5\%, 6\% and 15\%.

Uncertainties in the signal prediction from the scalar and vector DM mediator productions are also considered.  Missing higher-order QCD corrections are estimated by varying the renormalisation and factorisation scale by a factor of $0.5$ and $2$ relative to the value used in the baseline simulation. When the signal prediction is obtained with the reweighting procedure, two additional sources of uncertainty are considered. The first is a shape uncertainty, estimated by comparing the normalised \xgb~distributions at particle and reweighted level for the target point in the model parameter space.
The second is a 10\% normalisation uncertainty estimated as the largest difference between the event yields at particle and reconstruction levels for all the reference points.

Representative values for the impact of the considered sources of uncertainties in the background processes in the signal regions are shown in Table~\ref{tab:syst_table}. Generally, the dominant uncertainties in the background prediction consist of modelling uncertainties. The most important instrumental uncertainties are related to large-$R$ jet momentum and mass calibration, which significantly impacts the \vjets background expectation, and to top-tagging calibration, mostly affecting the \ttbar prediction.  %

%
%
%
%


%
%
%
%
%
%
%
%
%
%
%
%
%
%
%
%
%
%
%
%
%
%
%
%
%

%
%
%
%
%
%
%
%
%
%
%
%
%
%
%
%
%
%
%
%
%
%
%
%
%
%
%
%

%
%
%
%
%
%
%
%
%
%
%
%
%
%
%
%
%
%
%
%
%
%
%
%
%
%
%
%

%

\begin{table}[t]
\caption{
Values of the relative post-fit uncertainty (in \%) on the background prediction in the signal regions SR0b and SR1b for the different sources of systematic uncertainties.
The category ``jet calibration''  includes all the sources of uncertainty related to the reconstruction and calibration of small- and large-$R$ jets. The quoted values are obtained by averaging  the post-fit uncertainties of the three fits performed assuming the background-only hypothesis, corresponding to signal regions and BDTs targeting the scalar and vector mediator DM signals and the vector-like $T$ signal. The sum in quadrature of the individual uncertainties is not expected to be equal to the total background post-fit uncertainty computed taking into account the correlations among the sources of uncertainties. %
}
\begin{center}
\resizebox{\textwidth}{!}{
\begin{tabular}{@{}lcccccccccccc@{}}
\toprule
& \multicolumn{2}{c}{$t\bar{t}$} & \multicolumn{2}{c}{$Z${+jets}}  & \multicolumn{2}{c}{$W${+jets}}& \multicolumn{2}{c}{$tW$}  & \multicolumn{2}{c}{Diboson} & \multicolumn{2}{c}{$t\bar{t}V$}  \\
& {\footnotesize SR0b} & {\small SR1b} & {\footnotesize SR0b} & {\footnotesize SR1b}  & {\small SR0b} & {\footnotesize SR1b}  & {\footnotesize SR0b} & {\footnotesize SR1b}  & {\footnotesize SR0b} & {\footnotesize SR1b}  & {\footnotesize SR0b} & {\footnotesize SR1b} \\
\midrule
$b$-tagging efficiency & $5$ & $<1$  &$<1$  &$2$  &$<1$  &$2$ &$5$ & $<1$ &$<1$ &$3$ &$6$ &$1$  \\
Jet calibration  & $50$ & $7$ &$15$ & $30$ &$18$ & $13$ &$50$ & $13$ &$13$ & $20$ &$40$ & $12$\\
\met calibration & $3$ & $<1$ &$2$ & $6$ &$1$ & $<1$ &$2$ & $1$ &$<1$ & $2$ &$4$ & $1$ \\
Top-tagging efficiency & $13$ & $14$ &$5$ & $6$ &$5$ & $6$ &$13$ & $14$ &$11$ & $11$ & $13$ & $13$ \\
Other instrumental  & $3$ & $2$ &$2$ & $2$ &$3$ & $3$ &$5$ & $3$ &$3$ & $4$ &$5$ & $3$\\
Modelling & $30$ & $23$ & $23$ & $33$ &$20$ & $30$ & $65$ & $45$& $10$ & $35$ & $15$ & $15$ \\
\bottomrule
\end{tabular}
}
\label{tab:syst_table}
\end{center}
\end{table}



%

%
\FloatBarrier
\section{Results}
\label{sec:results}

For each signal model, the presence of the signal in data is tested via a profile-likelihood fit to the observed event yields in the control regions and to the corresponding \xgb~score distribution in the signal regions. The binning is chosen as a compromise between the maximum separation of the signal from the background and statistics of simulated background events.
The likelihood function used to construct the test statistic consists of a product of Poisson probabilities of the yields in each bin of the fitted distributions and constraint functions for NPs describing the systematic uncertainties.
The impact of statistical uncertaities from the data and MC simulation are parametrised by a Poisson distribution while the remaining uncertainties are parametrised by Gaussian distributions.
Additionally, the normalisation factors of \vjets (\muvjets) and \ttbar (\muttbar) background processes are included as unconstrained NPs.

Only systematic uncertainties impacting the shape or the normalisation of the expected background distribution  for more than $1\%$ are accounted for in the systematic model. In case a systematic variation leads to significant bin-to-bin statistical fluctuations in the \xgb~score distribution in the signal regions, the shape of the systematic effect is smoothed. Systematic uncertainties that are estimated via a direct comparison of a single variation with the nominal prediction are symmetrised.
Some systematic uncertainties that are evaluated via the comparison of two opposite variations with respect to the nominal prediction,  may show both variations above or below the nominal expectation with at least one variation compatible with the nominal prediction within the associated statistical uncertainty. In this case, the maximum distance from the nominal prediction is considered and symmetrised.
A test statistic based on a profile-likelihood ratio implemented in \textsc{RooStats}~\cite{RooFit} is used for the evaluation of confidence intervals and hypothesis testing. Exclusion limits are derived with the CL$_\text{s}$ frequentist method~\cite{Junk:1999kv,Read:2002hq} using the asymptotic approximation~\cite{Cowan:2010js}.

\begin{figure}[t]
\centering
\includegraphics[width=0.8\textwidth]{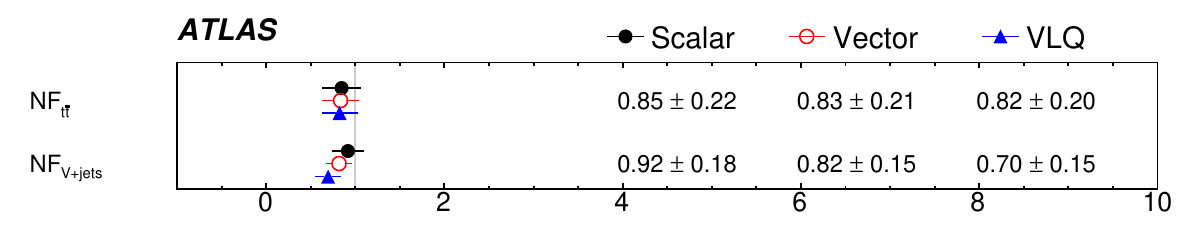}
\caption{
Comparison of the fitted normalisation factors of the \ttbar and $V${+jets} background processes in the fits of data in control and signal regions under the background-only hypothesis. An independent fit is performed for each signal model, using the respective signal regions.}
\label{fig:nf_comp}
\end{figure}

\begin{figure}[t]
\centering
\subfloat[]{\includegraphics[width=0.4\textwidth]{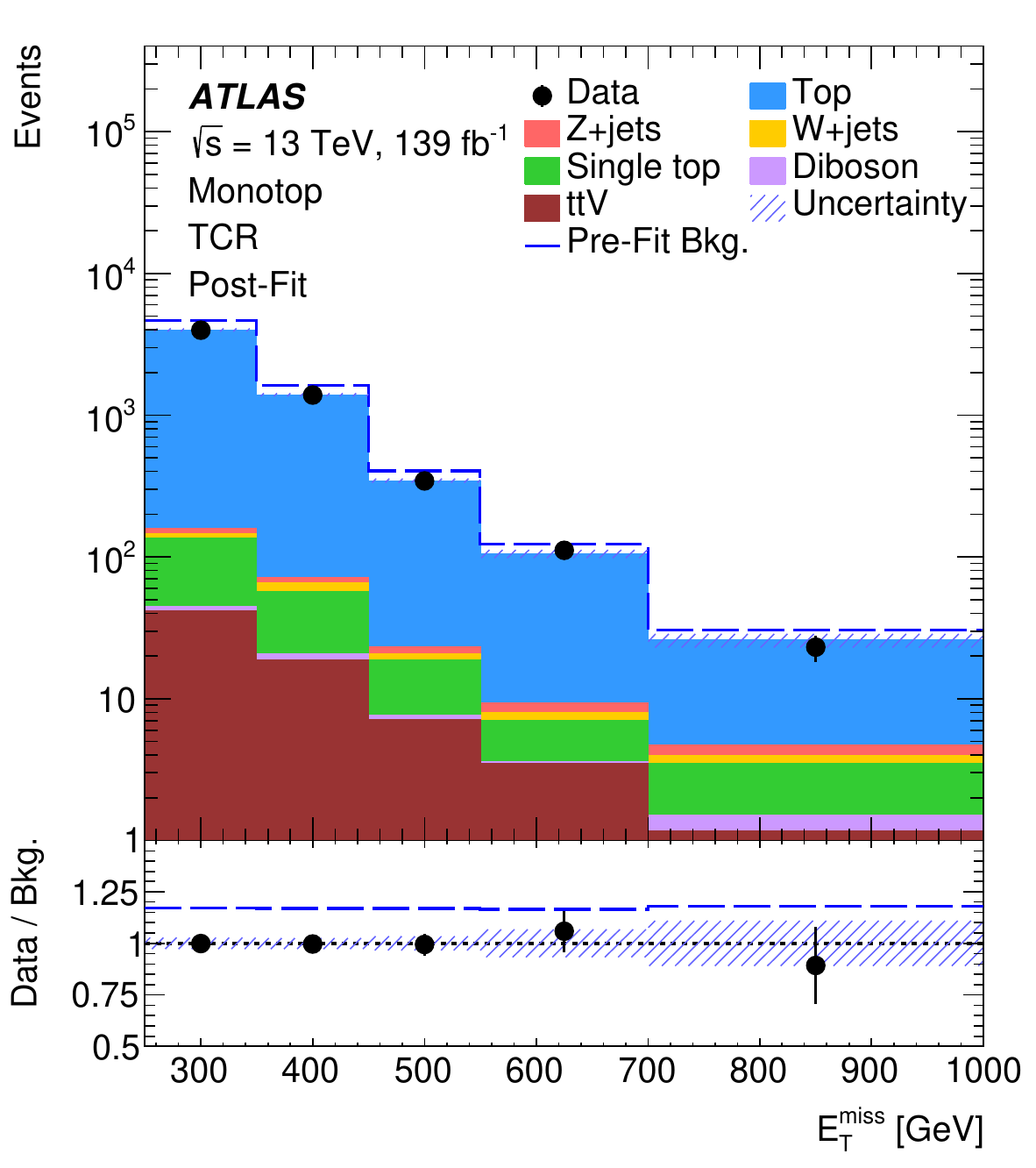}}
\subfloat[]{\includegraphics[width=0.4\textwidth]{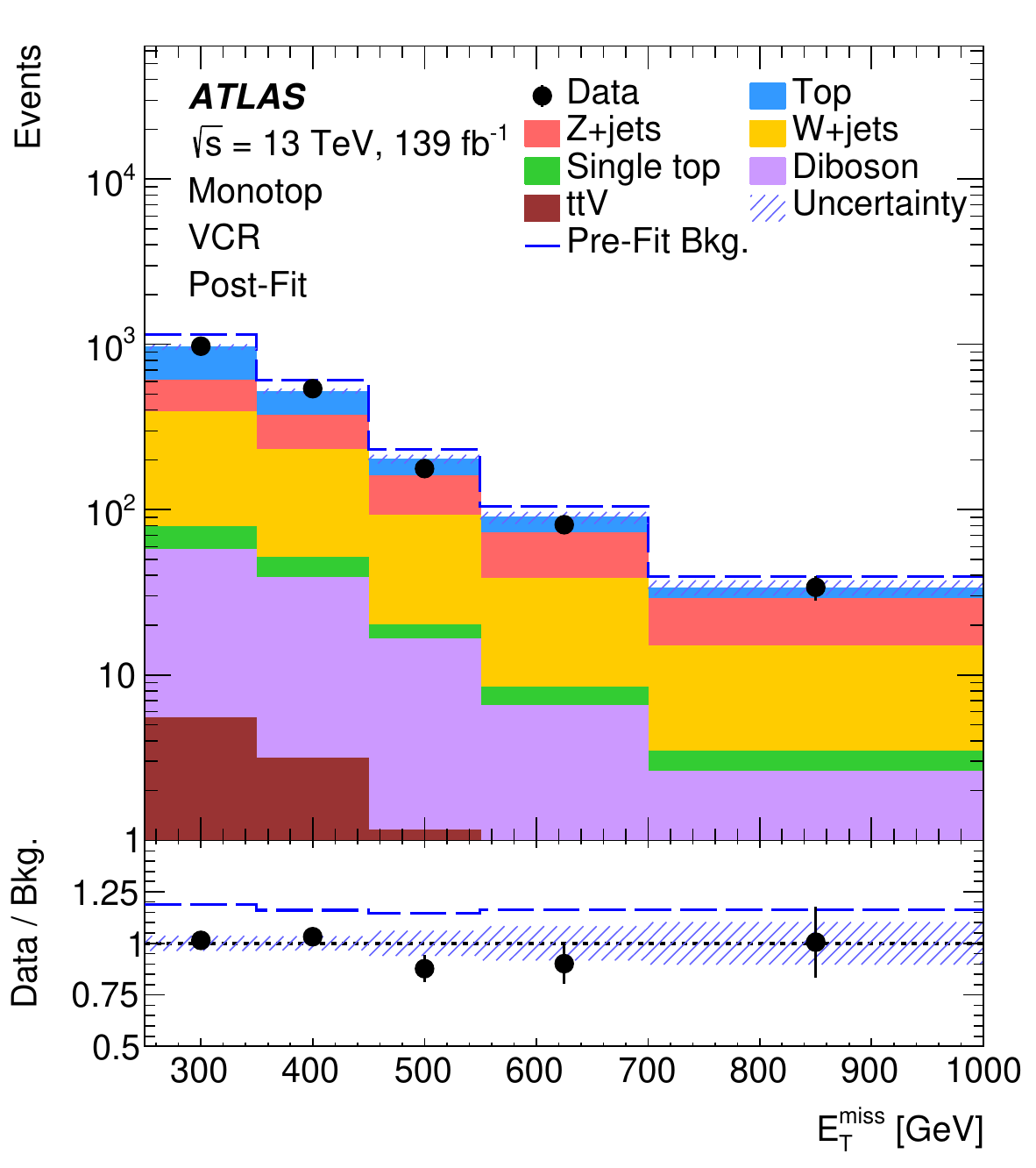}}
\caption{Comparison of data and SM prediction for the $\met$ distribution in control regions targeting (a) \ttbar and (b) \vjets processes. The background predictions with the corresponding uncertainties result from the simultaneous fit to data in control and signal regions under the background-only hypothesis (Post-Fit). The signal regions included in the fit correspond to those targeting the scalar DM mediator production. The dashed line indicates the pre-fit total background prediction. The last bin of the distribution contains the overflow.}
\label{fig:cr}
\end{figure}

The result of the fit under the background-only hypothesis leads to the determination of the normalisation factors for the main sources of background. The values obtained by the different fits are shown in Figure~\ref{fig:nf_comp}.
They are consistent with unity within two standard deviations.
The discrepancy observed in \muvjets~between the fits targeting DM and vector-like $T$ signals is explained by the different phase spaces of the corresponding signal regions (in particular due to the requirement of at least one forward jet in the VLQ signal regions).
The deviation from the unity of the normalisation factors is mainly related to the highly boosted regime explored in this analysis with respect to the phase space in which the top-quark tagging calibration was performed.
The good description of data in the control regions of the post-fit background model is shown in Figure~\ref{fig:cr} for the \met distribution. Despite that only the event yield information is used as fit input for the CRs, a very good modelling of the \met distribution is observed.

\begin{figure}[t]
\centering
\subfloat[]{\includegraphics[width=0.3\textwidth]{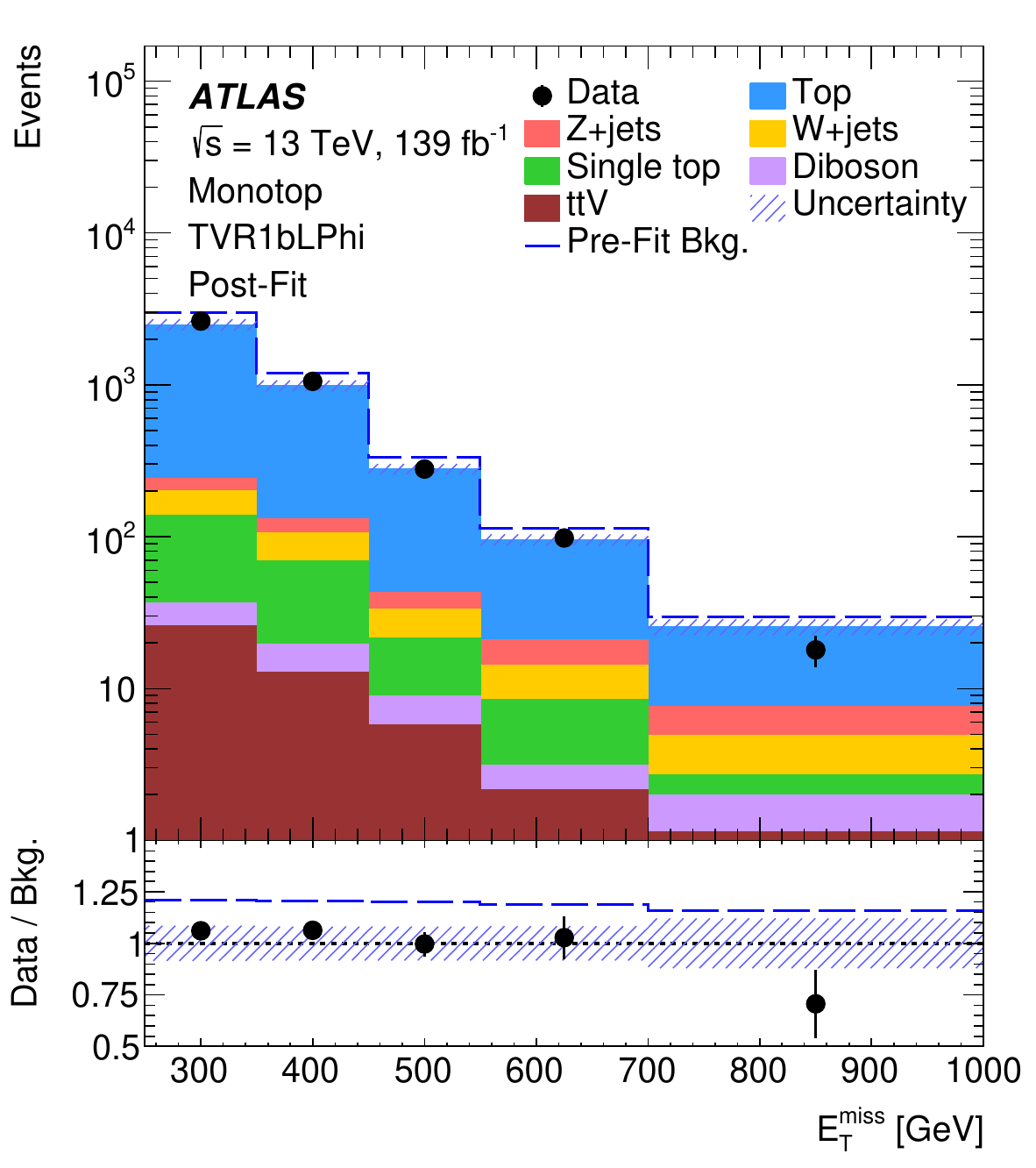}\label{fig:ttvr_res}}
\subfloat[]{\includegraphics[width=0.3\textwidth]{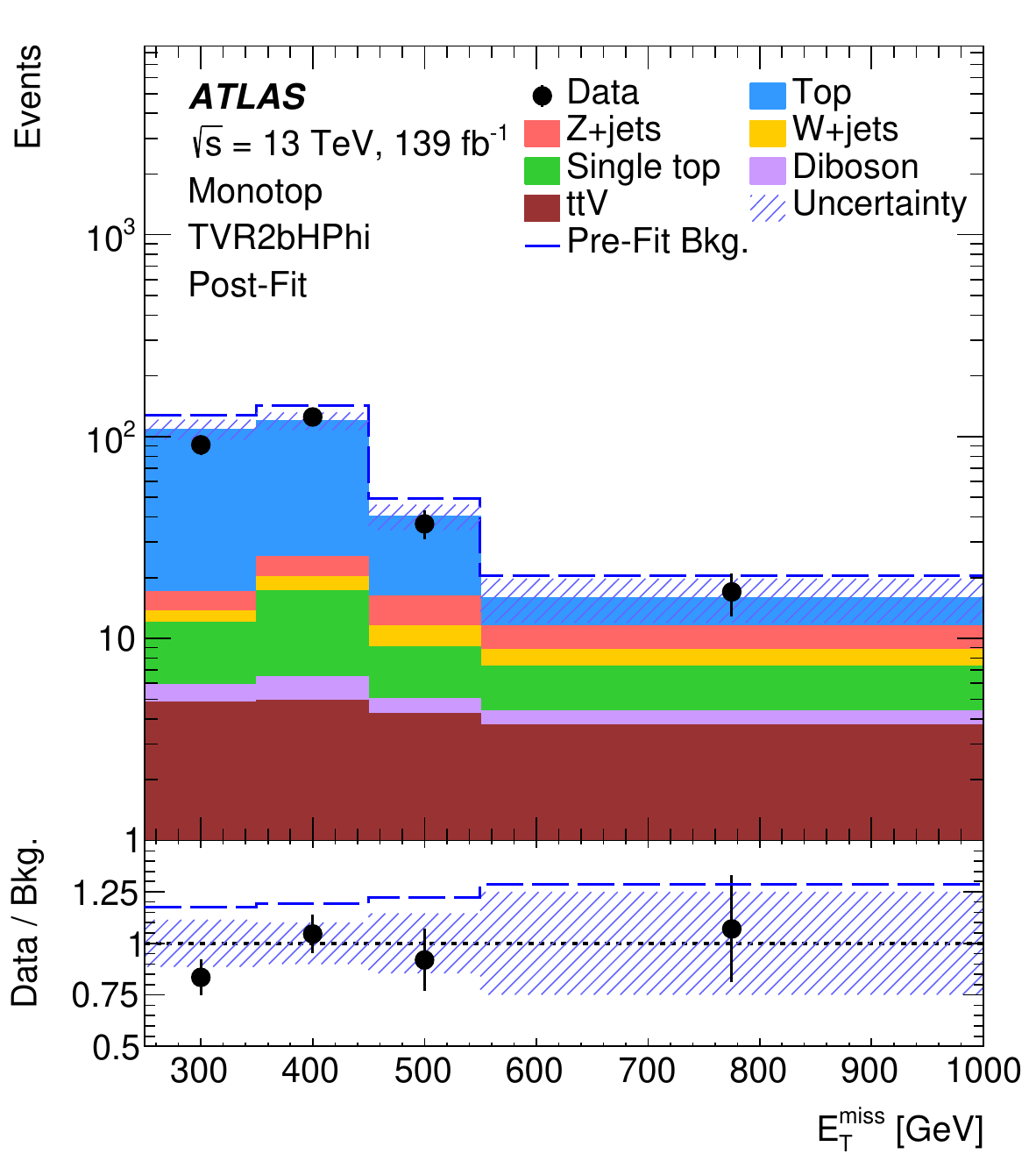}\label{fig:ttvr2_res}} \\
\subfloat[]{\includegraphics[width=0.3\textwidth]{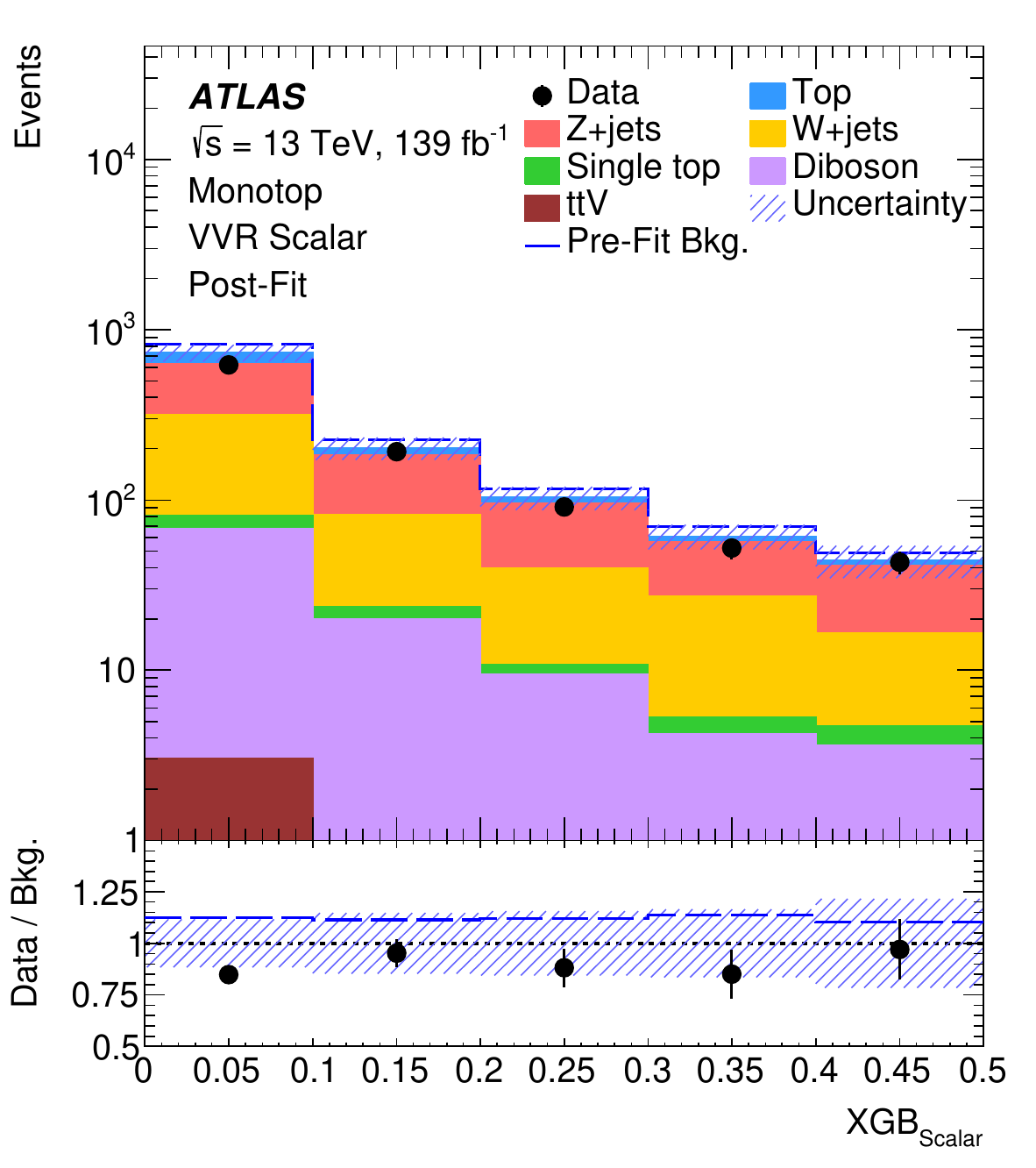}\label{fig:vvr0b_res}}
\subfloat[]{\includegraphics[width=0.3\textwidth]{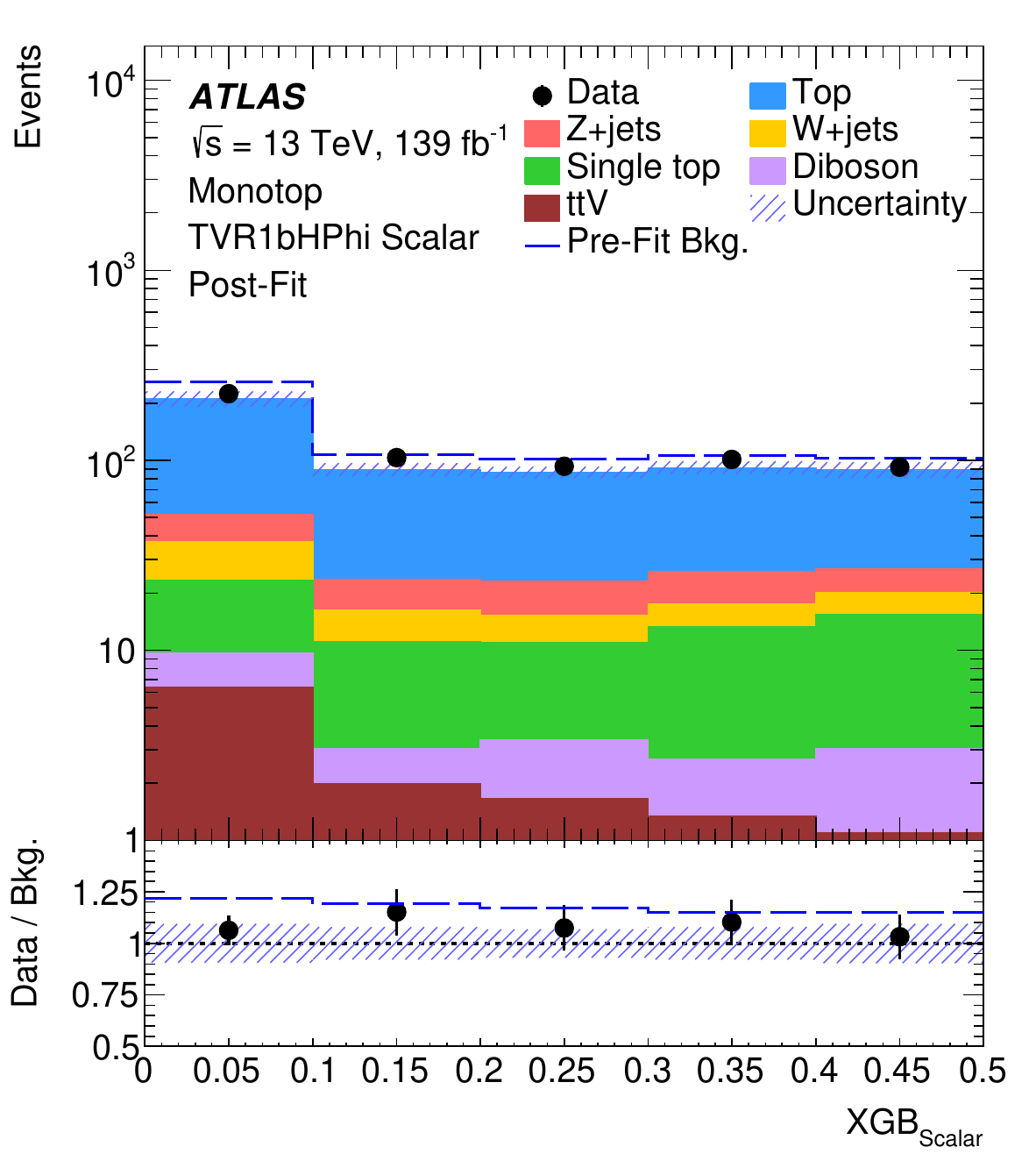}\label{fig:vvr1b_res}}
\caption{Comparison of data and SM prediction in the validation regions. The \met distribution is shown in the \ttbar validation regions  (a) TVR1bLPhi and (b) TVR2bHPhi. The \xgb~score is shown in the validation regions (c) VVR and (d) TVR1bHPhi, which are inclusive in forward jet multiplicity. The background predictions with the corresponding uncertainties result from the simultaneous fit to data in control and signal regions under the background-only hypothesis (Post-Fit). The signal regions included in the fit correspond to those targeting the scalar DM mediator production. The dashed line indicate the pre-fit total background prediction. The last bin of the distributions in (a) and (b) contains the overflow.}
\label{fig:vrs_res}
\end{figure}

\begin{figure}[t]
\centering
\subfloat[]{\includegraphics[width=0.45\textwidth]{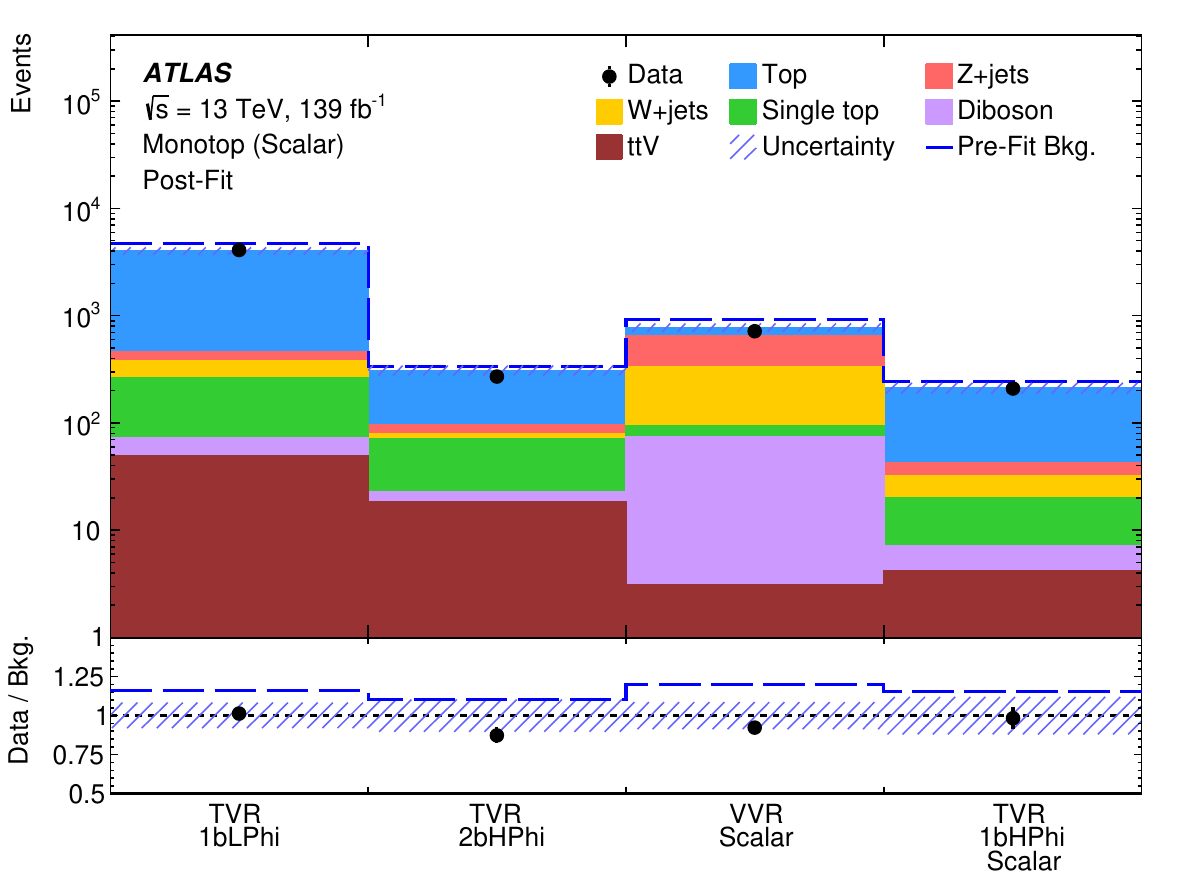}\label{fig:summary_vr_res}}
\subfloat[]{\includegraphics[width=0.45\textwidth]{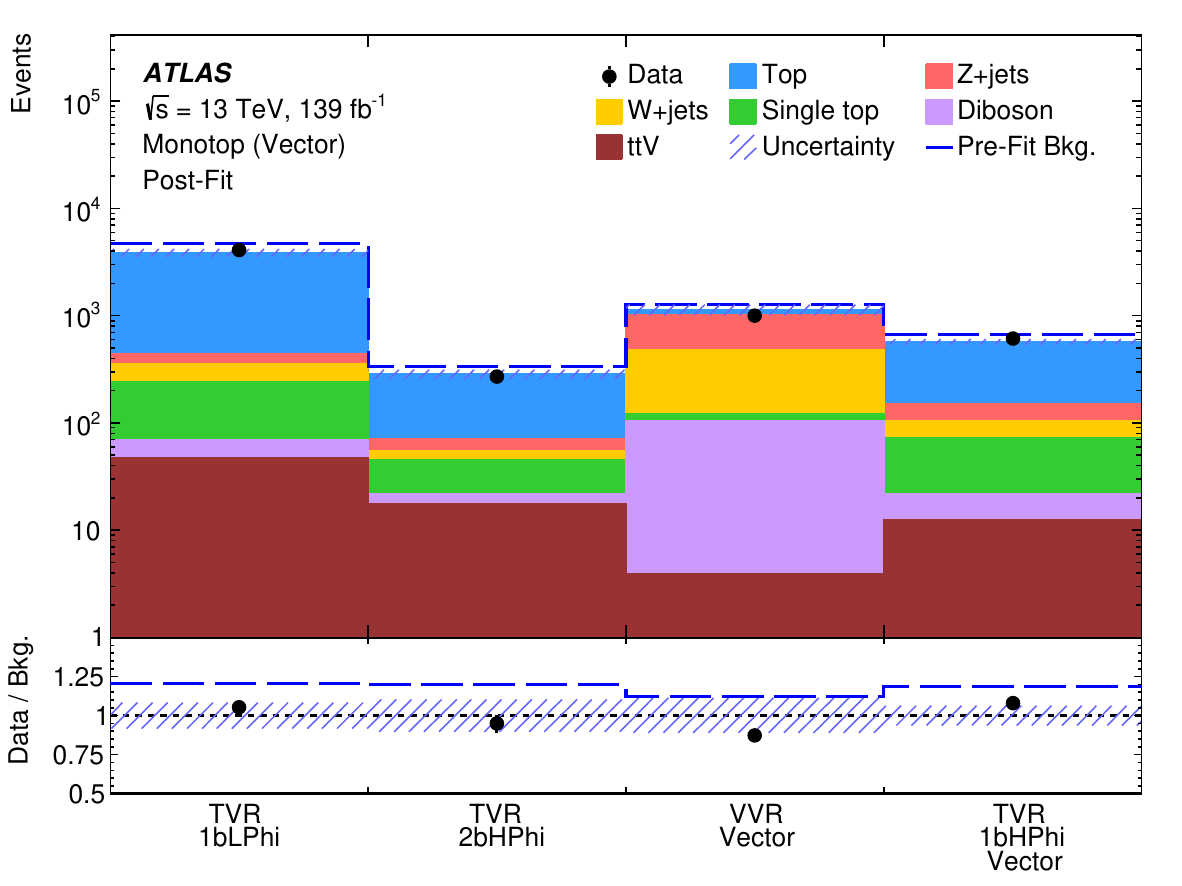}\label{fig:summary_vr_nonres}}\\
\subfloat[]{\includegraphics[width=0.45\textwidth]{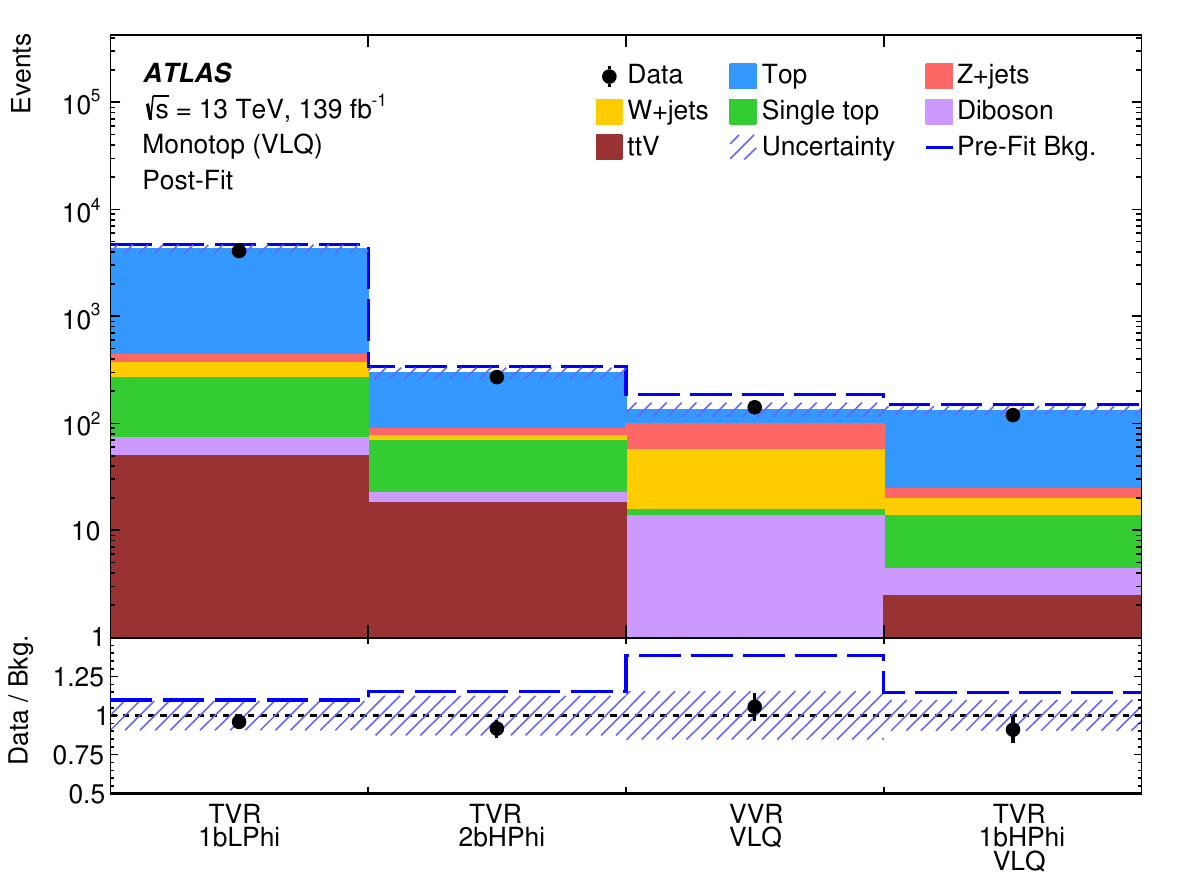}\label{fig:summary_vr_vlq}}
\caption{Comparison of data and SM predictions for the event yields in each validation region defined for the (a) DM scalar mediator, (b) DM vector mediator and (c) vector-like $T$ search. The background predictions with the corresponding uncertainties result from the simultaneous fit to data in control and signal regions under the background-only hypothesis (Post-Fit). The signal regions included in each fit correspond to those targeting the respective signal model.
The dashed line indicate the pre-fit total background prediction.}
\label{fig:summary_vr}
\end{figure}

The extrapolation of the background model from the control region to the signal region is tested in the validation regions.
These regions are not included in the likelihood fit but are used to validate the background expectation, obtained from the results of the fit in the control and signal regions, in regions kinematically close to the signal phase space.
Kinematical properties of \ttbar events are validated in adjacent regions to the \ttbar control region and common to all considered models, namely TVR1bLPhi and TVR2bHPhi, shown in Figures~\ref{fig:ttvr_res} and \ref{fig:ttvr2_res} for the \met distribution.
The validation regions that are closest to the signal regions are VVR and TVR1bHPhi, which only differ by the \xgb~score requirement with the SR0b and SR1b regions, respectively. A comparison of the observed and expected distribution for the \xgb~score variable is shown in Figures~\ref{fig:vvr0b_res} and \ref{fig:vvr1b_res}, respectively. No significant mismodelling of data is observed in these regions, meaning that the lower tail of the \xgb~score distribution in $N_{b\text{-tagged jets}}$--\dphimin~phase space is well modelled. The agreement of the event yields in data with the background model in all the validation regions considered in the analysis is shown in Figure~\ref{fig:summary_vr} for each signal model targeted.  Since no significant discrepancy between  data and the background expectation was observed in any of the validation regions, no additional non-closure  uncertainties are included in the systematic model of the profile-likelihood fit.

\begin{figure}[p]
\centering
\subfloat[]{\includegraphics[width=0.3\textwidth]{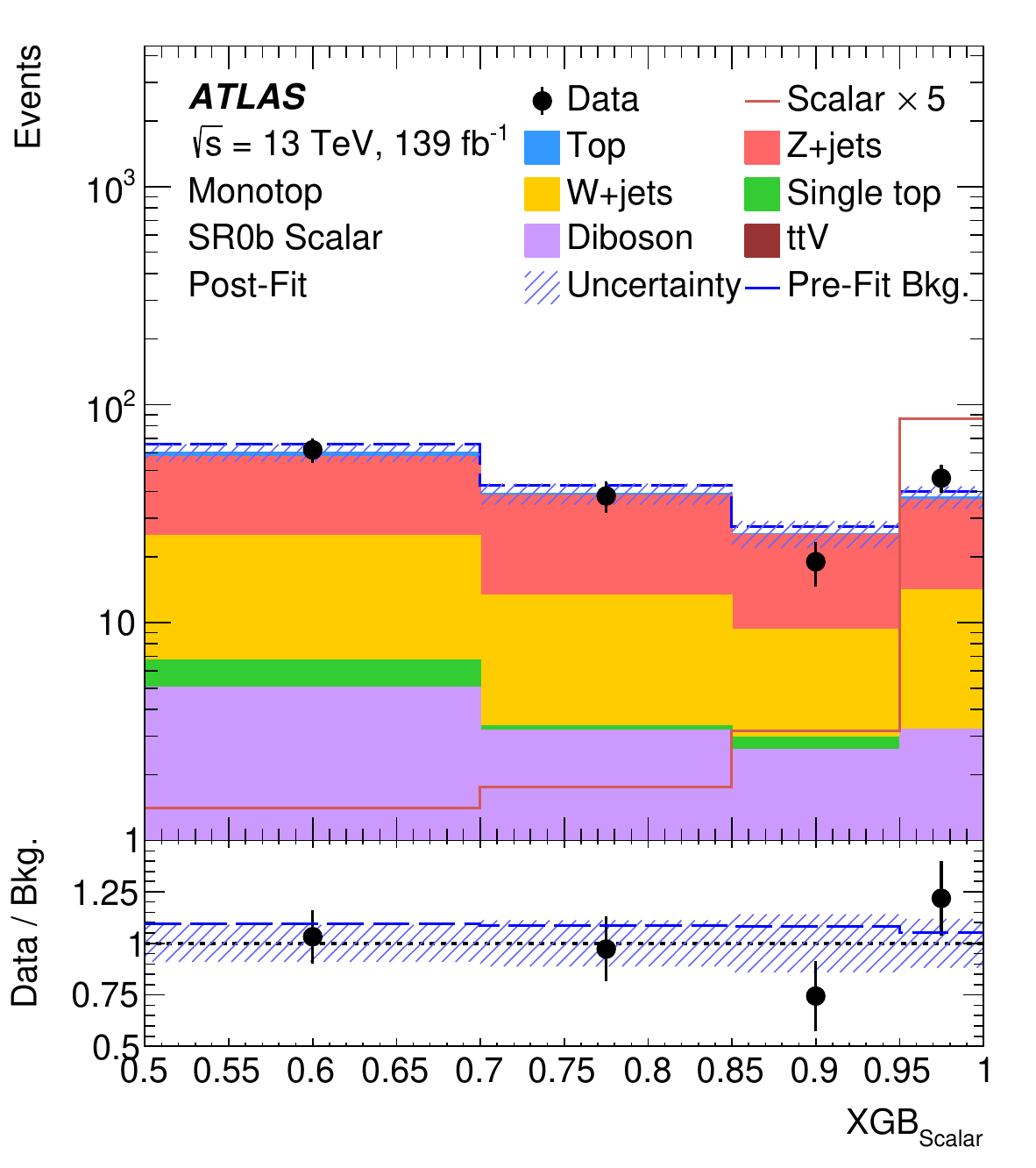}}
\subfloat[]{\includegraphics[width=0.3\textwidth]{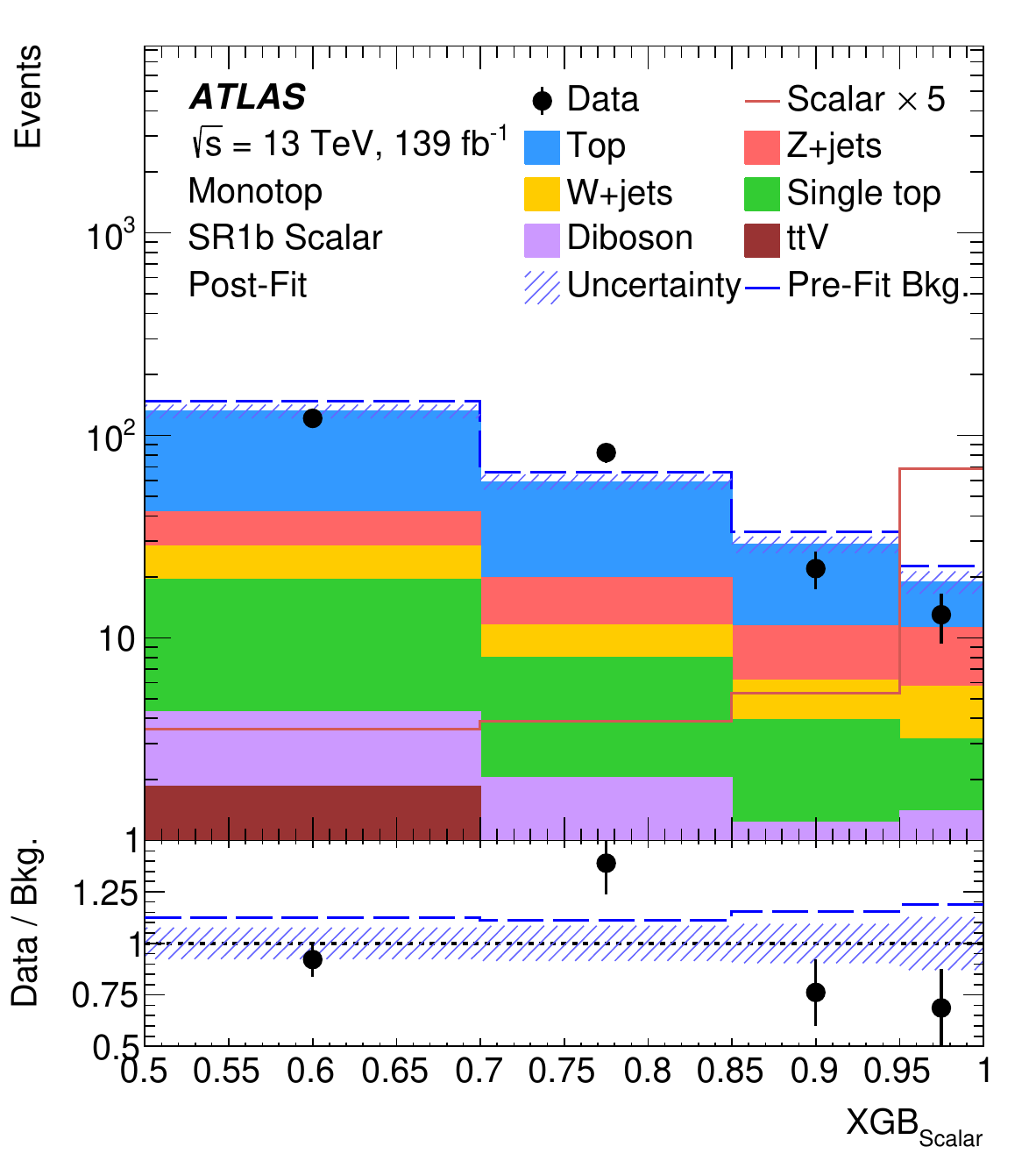}}\\
\subfloat[]{\includegraphics[width=0.3\textwidth]{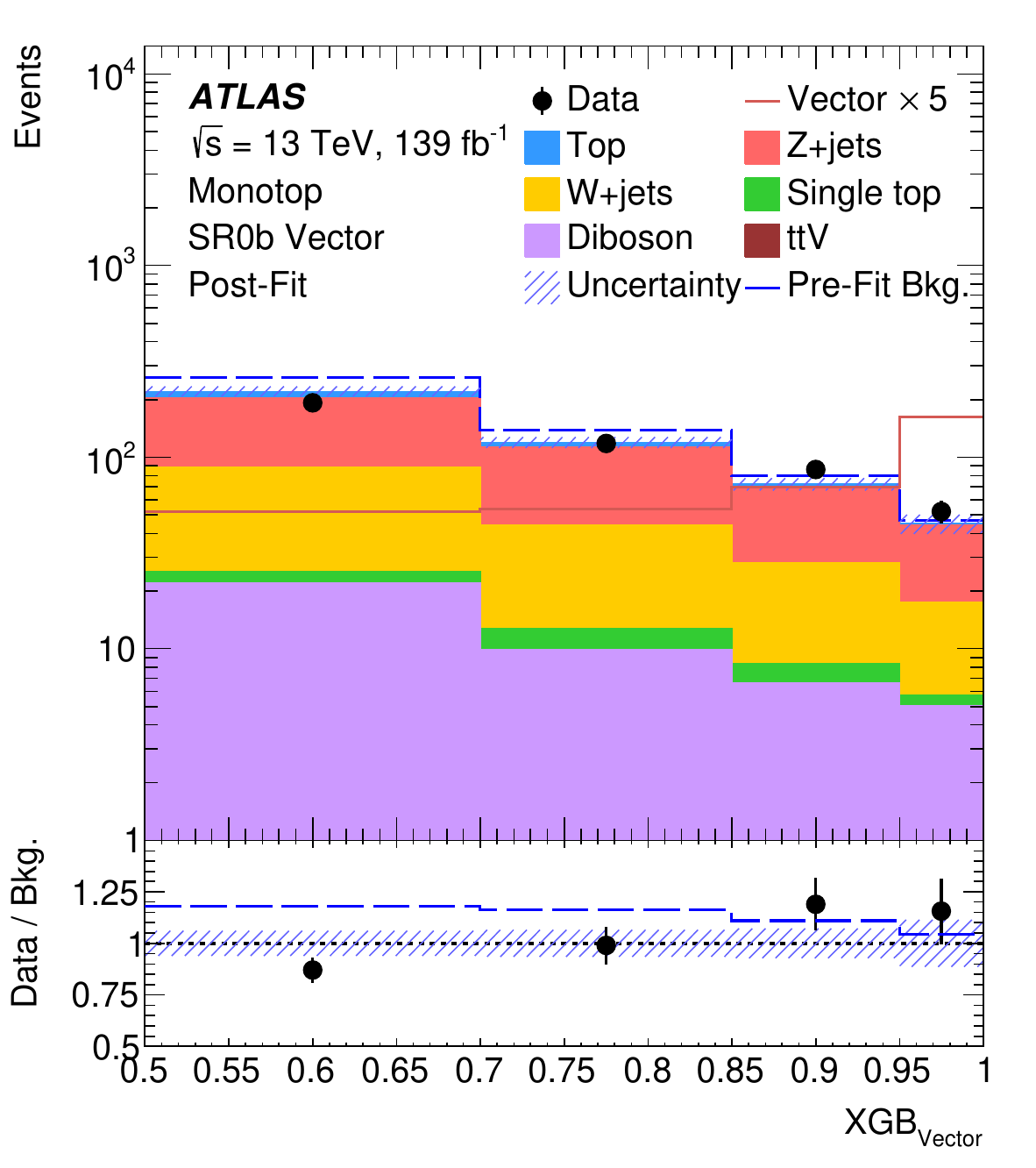}}
\subfloat[]{\includegraphics[width=0.3\textwidth]{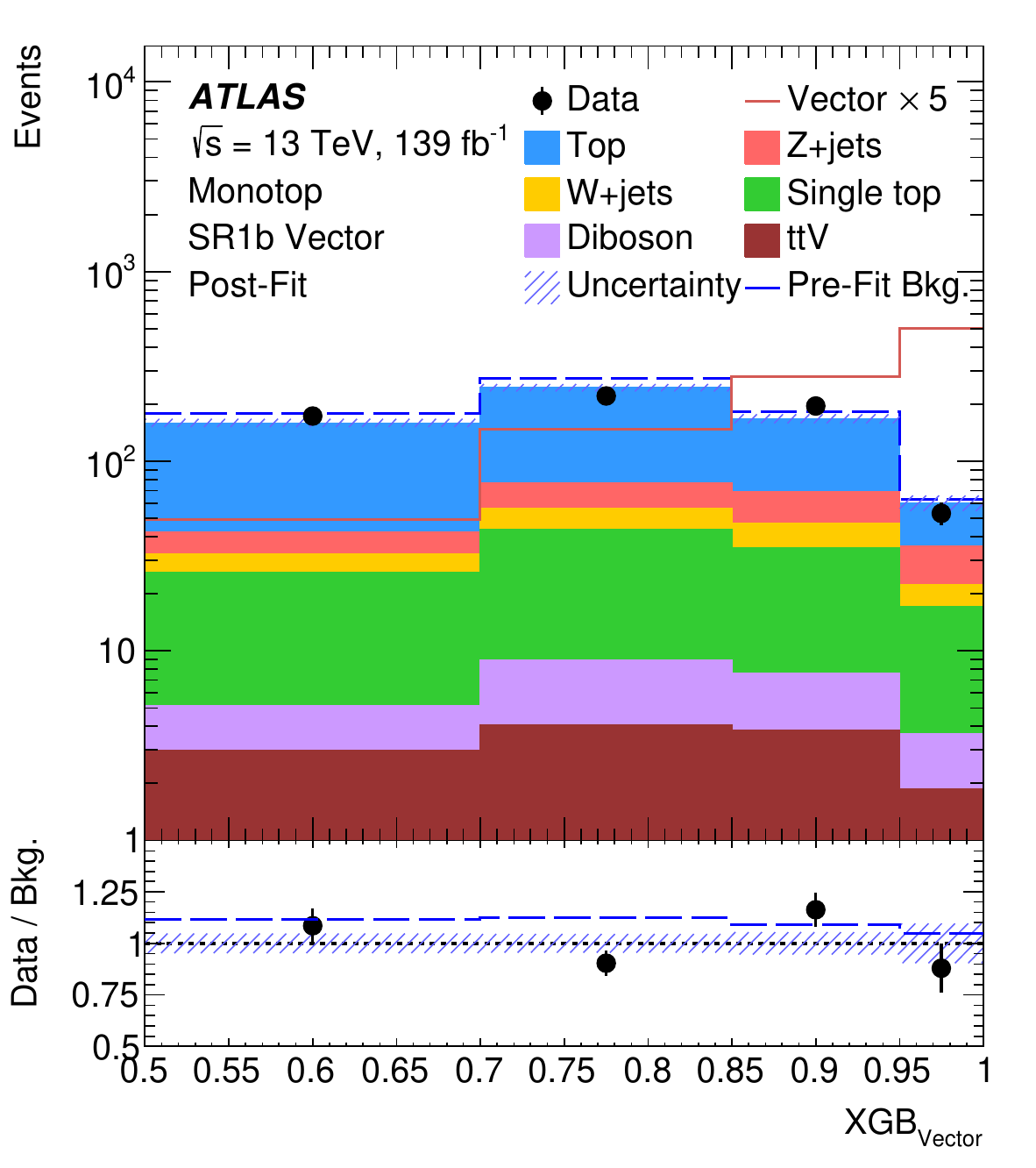}} \\
\subfloat[]{\includegraphics[width=0.3\textwidth]{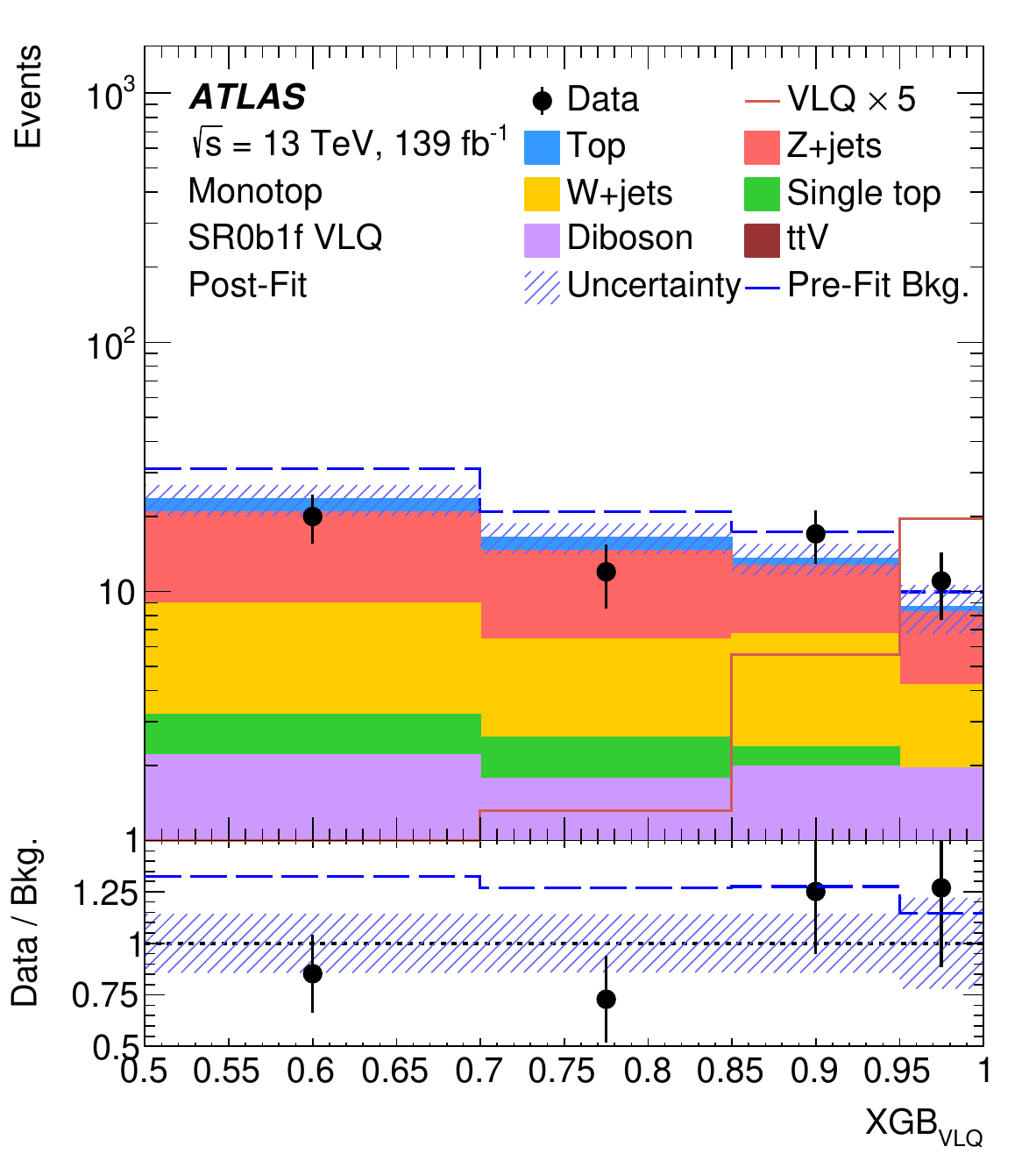}}
\subfloat[]{\includegraphics[width=0.3\textwidth]{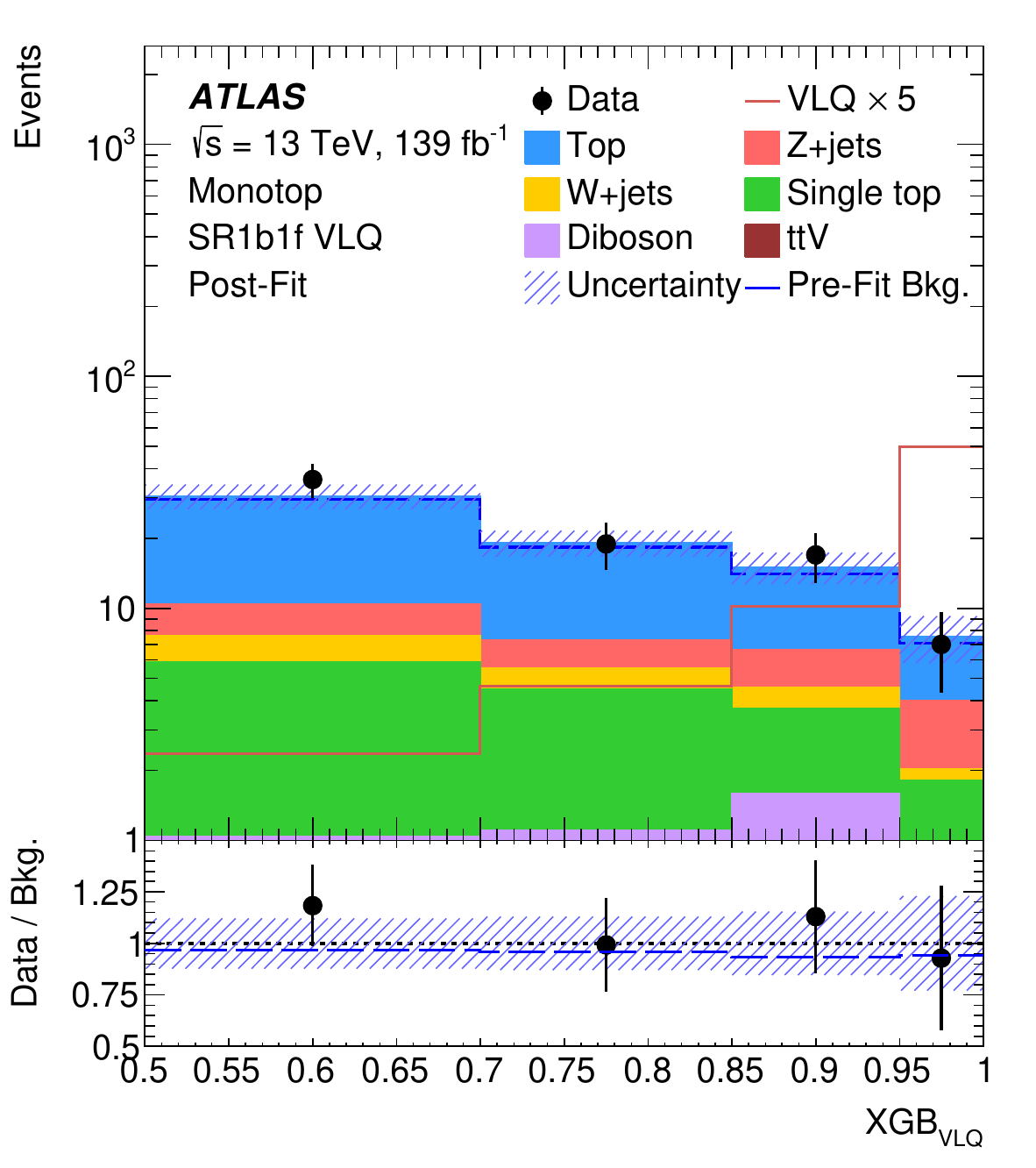}}
\caption{Comparison of data and fitted expectation for the \xgb~score distribution in the signal regions targeting the three considered signal models: (a) SR0b and (b) SR1b for the scalar DM mediator production, (c) SR0b and (d) SR1b for the vector DM mediator production and (d) SR0b (e) SR1b for the single vector-like $T$ quark production. The background predictions with the corresponding uncertainties result from the simultaneous fit to data in control and signal regions under the background-only hypothesis (Post-Fit). The signal regions included in each fit correspond to those targeting the respective signal model. The dashed line indicates the pre-fit total background prediction. The overlayed signal distributions correspond to scalar DM mediator production with $m_\phi=4$~\TeV (Scalar), vector DM mediator production with $m_V =1.75$~\TeV (Vector) and single vector-like $T$ production with $m_T=1.7$~\TeV (VLQ) scaled by a factor $5$.}
\label{fig:sr}
\end{figure}

The distribution of the \xgb~score in the signal regions for data and the fitted SM expectation under the background-only hypothesis are shown in Figure~\ref{fig:sr}.   No significant excess above the SM expectation is found in any of the signal regions.
The results are therefore interpreted in terms of expected and observed upper limits on the signal cross-section as a function of the model parameters.

\begin{figure}[t]
\centering
\subfloat[]{\includegraphics[width=0.4\textwidth]{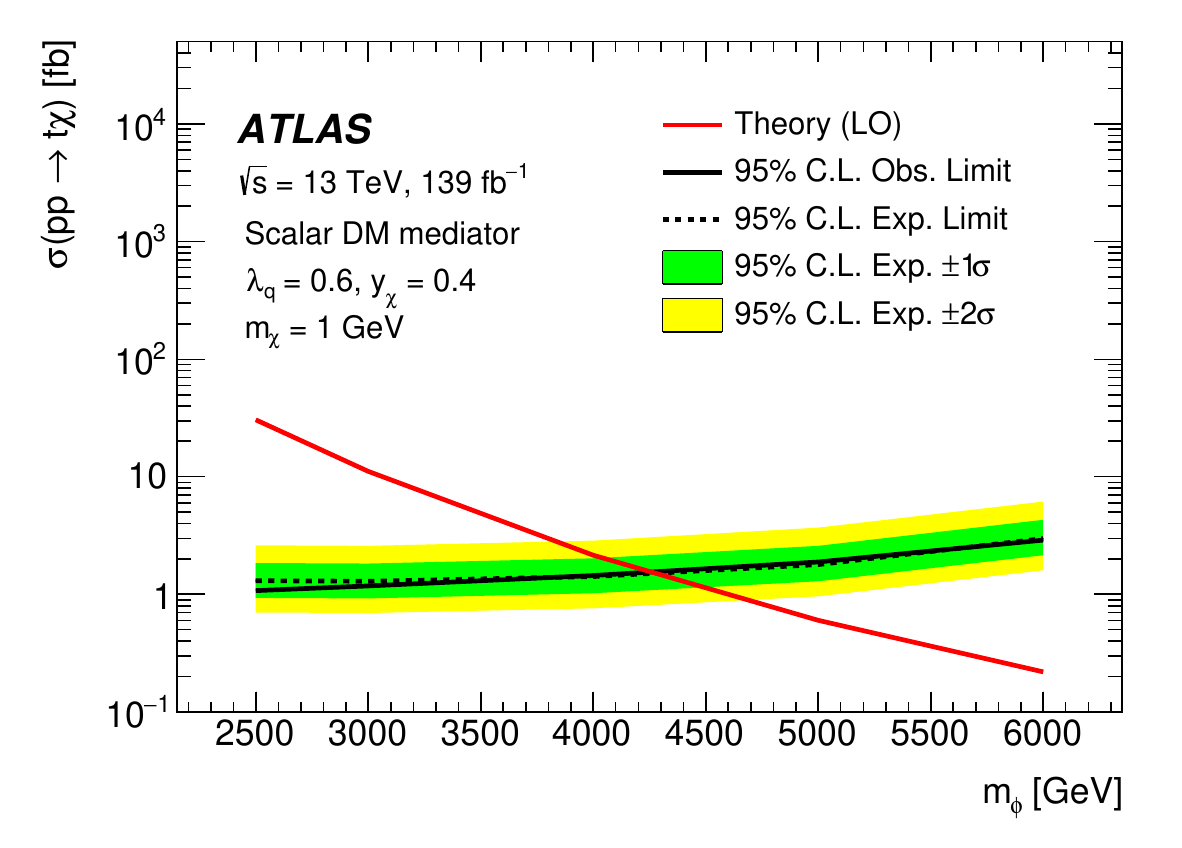}\label{fig:limit_res}}
\subfloat[]{\includegraphics[width=0.4\textwidth]{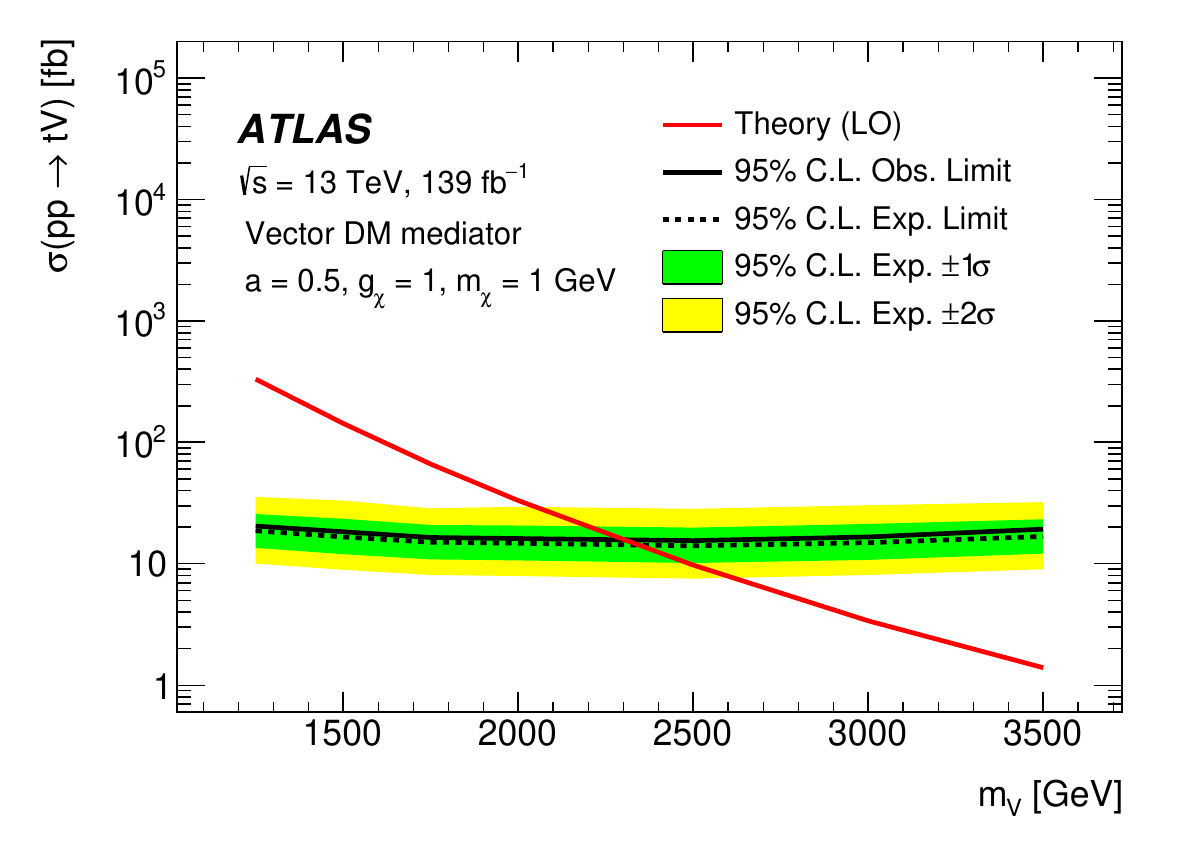}\label{fig:limit_nonres}}\\
\subfloat[]{\includegraphics[width=0.4\textwidth]{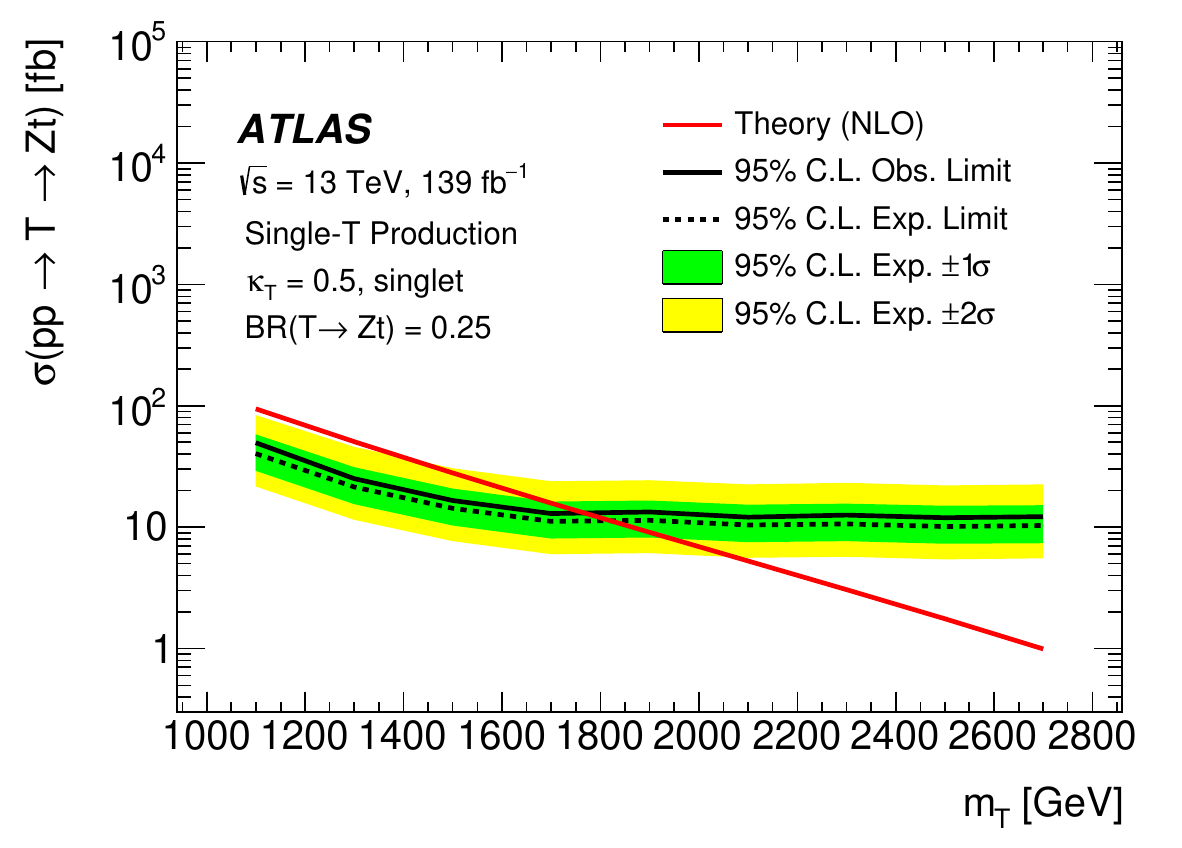}\label{fig:limit_vlq}}
\caption{ 95\% CL upper limits on the cross-section of the considered signal models as a function of: (a) the DM scalar mediator $\phi$ mass (for fixed model parameters of $\lambda_q =0.6$, $y_\chi=0.4$ and $m_\chi=1$~\GeV), (b) the DM vector mediator $V$ mass (for $a=0.5$, $g_\chi=1$ and $m_\chi=1$~\GeV) and (c) the vector-like $T$ quark mass (for $\kappa_T=0.5$).}

\label{fig:limit}
\end{figure}

The upper limit on the signal cross-sections at 95\% confidence level is shown in Figure~\ref{fig:limit}, together with the theoretical signal cross-section prediction for the considered benchmarks. The limit on the cross-section of the scalar-mediated DM production as a function of the mediator mass $m_\phi$ is shown in Figure~\ref{fig:limit_res}. A degradation of the cross-section limit at high $m_\phi$ masses is observed, justified by a general decrease in the event selection efficiency. Assuming the couplings $\lambda_q=0.6$ and $y_\chi=0.4$, and the mass of the DM candidate $m_\chi=1$~\GeV, the model is excluded for masses of the mediator scalar particle $m_\phi<4.3$~\TeV, improving the limits with respect to the previous search with \emph{mono-top} events by $800$~\GeV~\cite{EXOT-2017-16}.
The limit on the cross-section of the vector-mediated DM production as a function of the mass of the vector particle $m_V$ is shown in Figure~\ref{fig:limit_nonres}. The predictions of the signal distribution for a mass of the vector mediator $m_V$ different from $1.75$~\TeV are obtained using the weighting procedure described in Section~\ref{sec:samples}. Assuming the couplings $a=0.5$ and $g_X=1$, and the mass of the DM candidate $m_\chi=1$~\GeV, the model is excluded for masses of the vector particle $m_V<2.3$~\TeV, which corresponds to an $300$~\GeV improvement from the previous result of the \emph{mono-top} channel~\cite{EXOT-2017-16}.

The limit on the cross-section of the vector-like $T$ quark production as a function of its mass $m_T$ is shown in Figure~\ref{fig:limit_vlq}. Assuming the coupling $\kappa_T=0.5$ and the branching ratio ${\mathcal{B}}(T\to Zt) = 25$\%, the singlet model is excluded for $m_T<1.8$~\TeV. This constitutes an improvement of $400$~\GeV in the excluded mass when comparing to the previous similar analysis~\cite{EXOT-2017-16}. The cross-section limit is decreased by a factor 10, mainly due to improved object reconstruction algorithms, calibrations and event selection, the use of \xgb for enhanced signal sensitivity and increased data statistics.
This result
is competitive with
the 1--2~\TeV exclusion limits obtained analysing different final state events in searches for single vector-like $T$-quarks~\cite{
EXOT-2014-12,
EXOT-2014-13,
EXOT-2018-52,
EXOT-2020-01,
CMS-B2G-15-008,
CMS-B2G-16-001,
CMS-B2G-16-005,
CMS-B2G-16-013,
CMS-B2G-17-007,
CMS-B2G-17-018,
CMS-B2G-18-003},
as discussed in Section~\ref{sec:intro}.

\begin{figure}[p]
\centering{}
\subfloat[]{ \includegraphics[width=0.45\textwidth,]{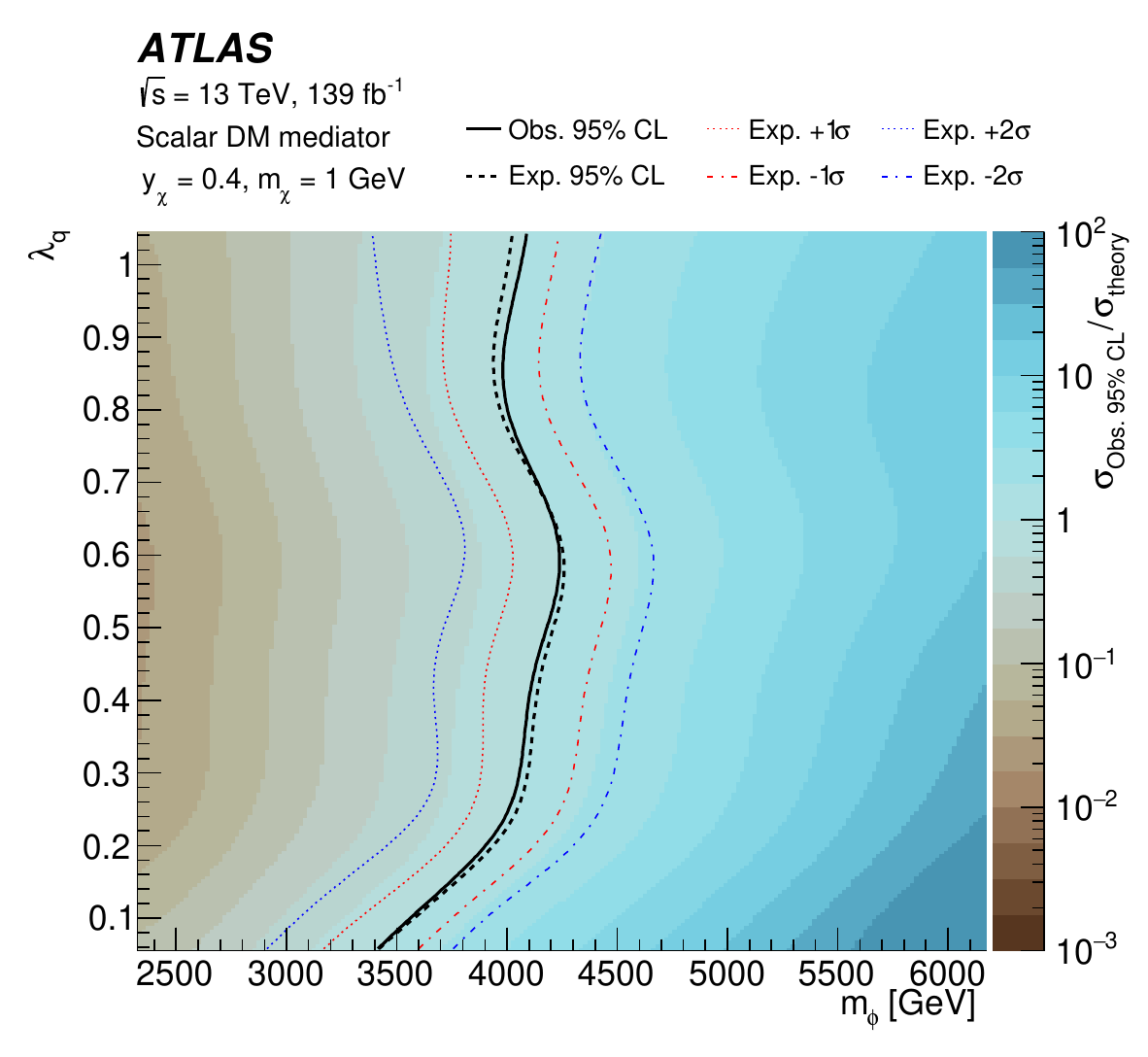}}
\subfloat[]{ \includegraphics[width=0.45\textwidth,]{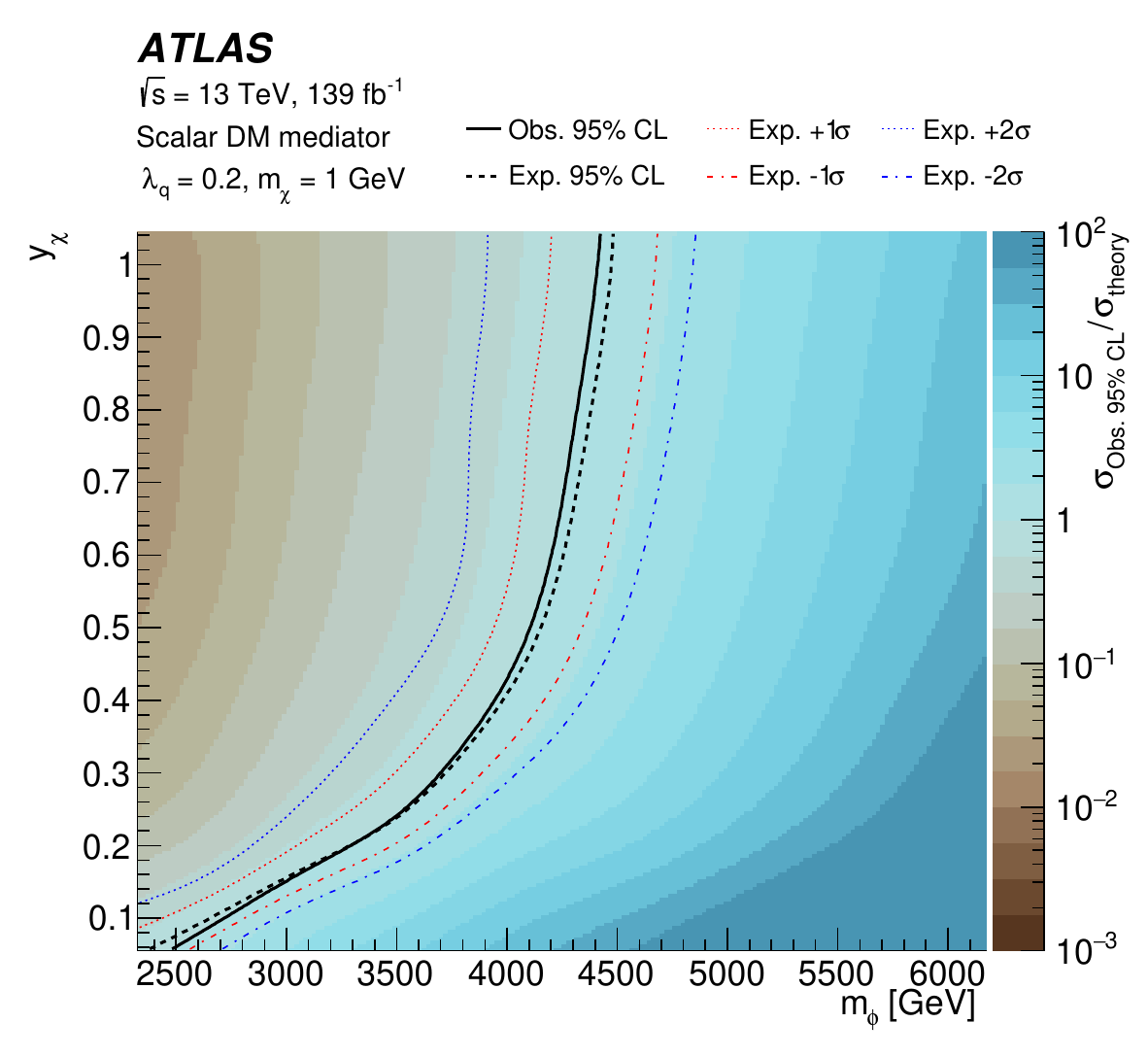}}\\
\subfloat[]{ \includegraphics[width=0.45\textwidth]{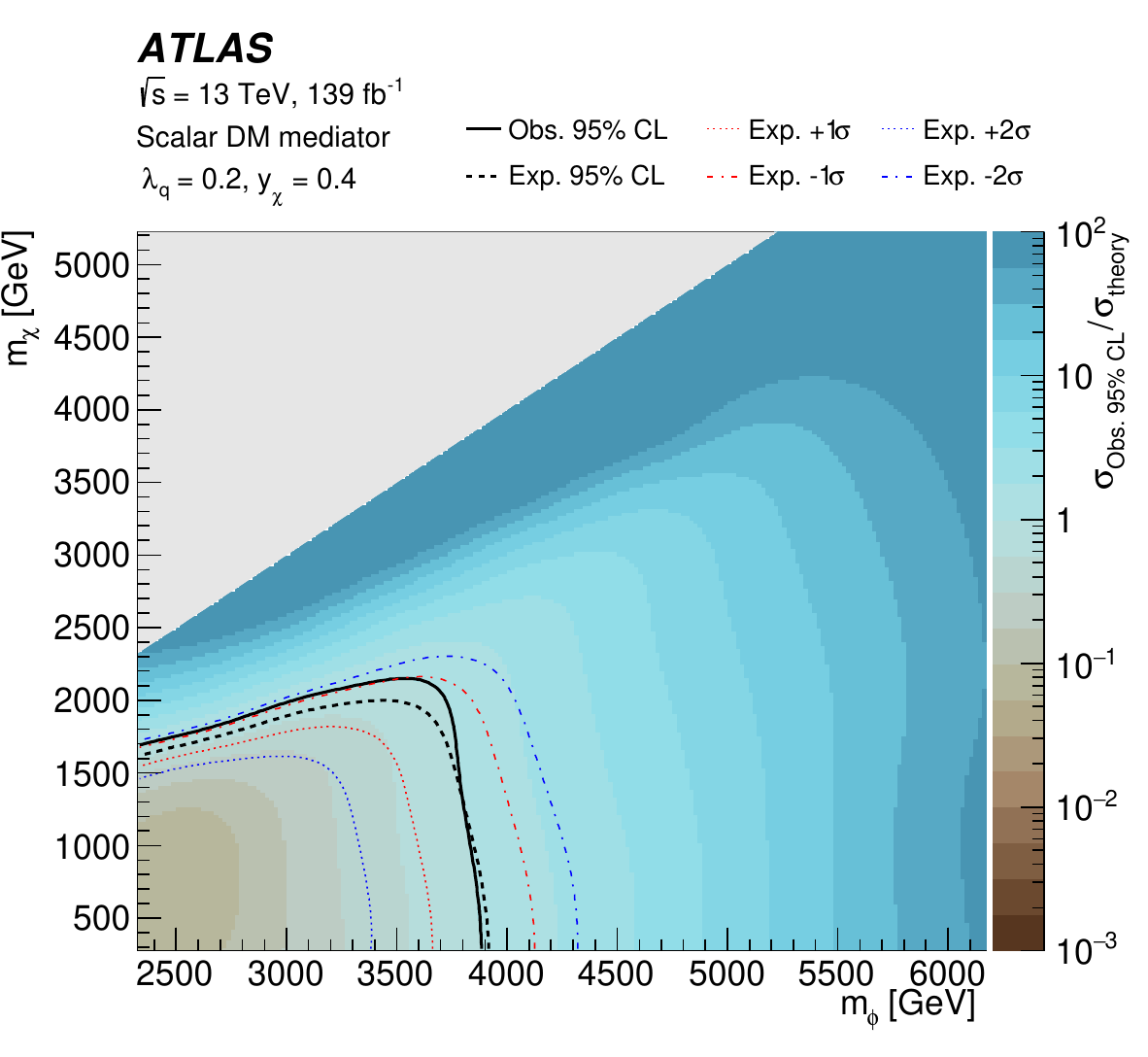}}
\caption{Observed 95\% CL upper limits on the scalar-mediated DM signal cross-section divided by the theoretical value in three planes of the model parameters space: (a) $(m_\phi, \lambda_q)$,  (b) $(m_\phi, y_\chi)$ and (c) $(m_\phi, m_\chi)$. The observed (expected) 95\% CL exclusion limits on the parameters are drawn as solid (dashed) lines. The $\pm1$ and $\pm2$ standard deviations around the expected limit are also shown.}
\label{fig:results_data_res_map2d}
\end{figure}

\begin{figure}[p]
\centering{}\subfloat[]{ \includegraphics[width=0.45\textwidth,]{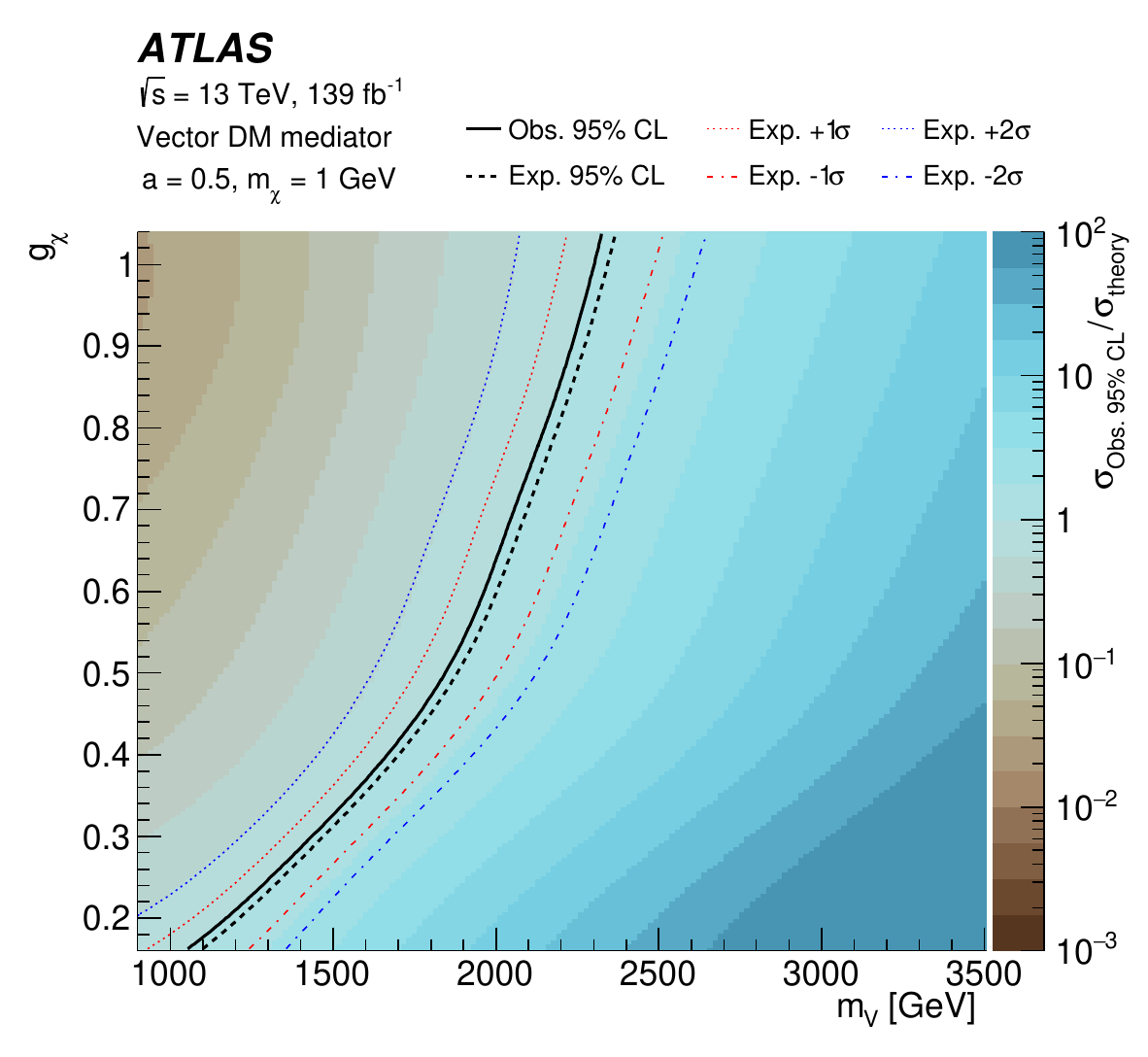}}
\subfloat[]{ \includegraphics[width=0.45\textwidth,]{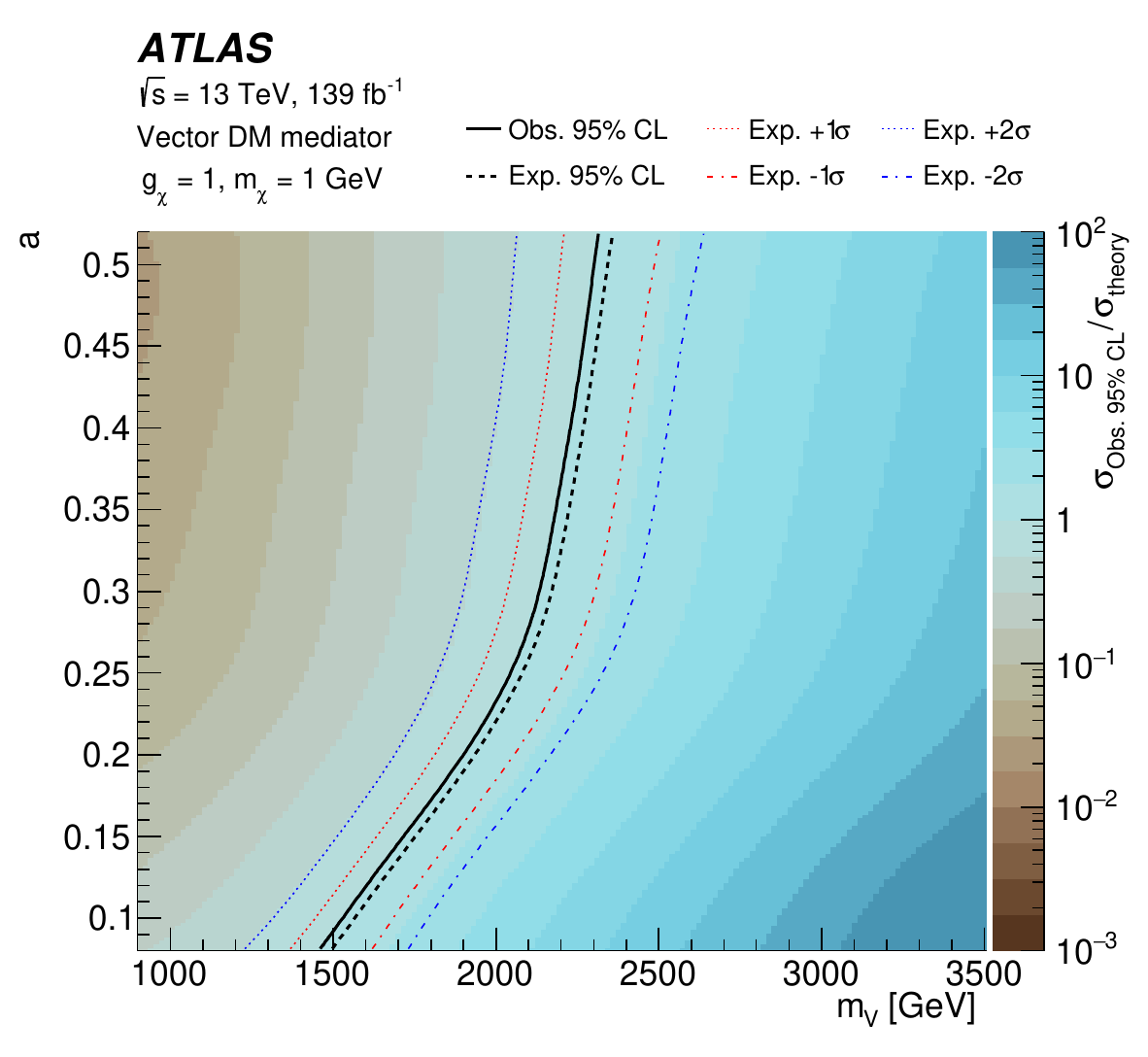}}\\
\subfloat[]{ \includegraphics[width=0.45\textwidth]{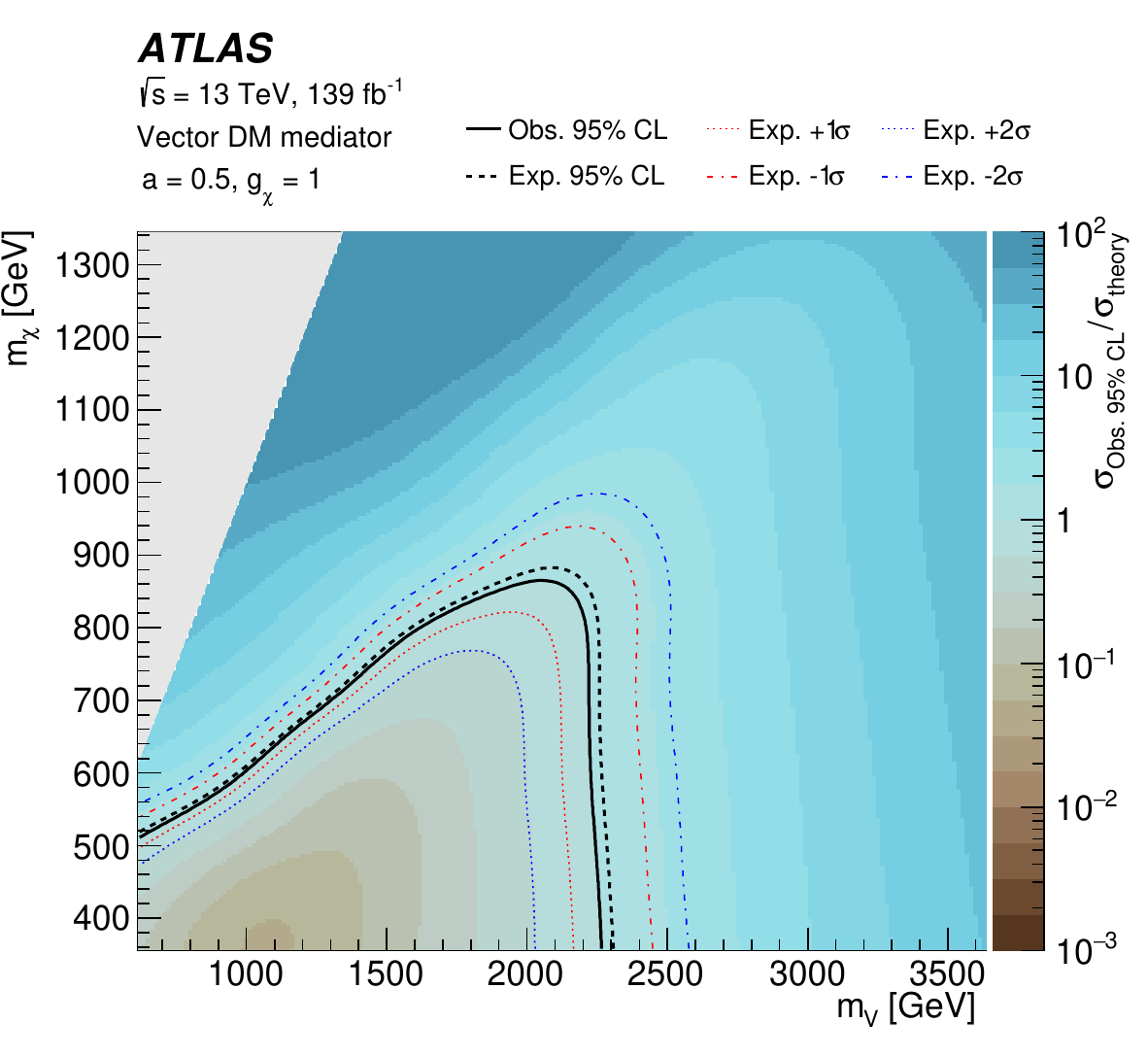}}
\caption{Observed 95\% CL upper limits on the vector-mediated DM signal cross-section divided by the theoretical value in three planes of the model parameters space: (a) $(m_V, g_\chi)$,  (b) $(m_V, a)$ and (c) $(m_V, m_\chi)$. The observed (expected) 95\% CL exclusion limits on the parameters are drawn as solid (dashed) lines. The $\pm1$ and $\pm2$ standard deviations around the expected limit are also shown.}
\label{fig:results_data_nonres_map2d}
\end{figure}

Contours of the observed and expected upper limits in four dimensions of the parameter space of the scalar and vector mediator DM models are produced using the set of simulations and reweighting procedure developed for this analysis. These are shown in Figures~\ref{fig:results_data_res_map2d} and \ref{fig:results_data_nonres_map2d}, respectively. The limit in the scalar mediator mass $m_\phi$ of the DM production appears to be almost constant for different values of $\lambda_q$ and $y_\chi$, except for very low values associated with tiny cross-sections, and all values of $\lambda_q$ ($y_\chi$) are excluded for $m_\phi$ up to $3.4$ ($2.5$)~\TeV, improving the results of the previous analysis. The $m_\phi$ limits are also constant in the $(m_\phi, m_\chi)$ plane, for the lower region of the DM candidate mass, and the $m_\chi$ reach is $2.2$~\TeV. This analysis is not sensitive to the region of the parameter space where $(m_\phi-m_\chi) < 500$~\GeV as it does not result in a boosted top-quark topology. For the vector DM mediator production model, the exclusion limits obtained are improved with respect to the previous analysis, all values of $g_\chi$ ($a$) are excluded for $m_V$ up to $1.1$ ($1.5$)~\TeV, and the reach in $m_\chi$ is improved by approximately $100$~\GeV. These results are complementary to the ones obtained in a search for same-charge top-quark pair production~\cite{EXOT-2016-16},
providing sensitivity to lower values of the $a$ coupling parameter.

\begin{figure}[t]
\centering{}
\subfloat[]{ \includegraphics[width=0.45\textwidth,]{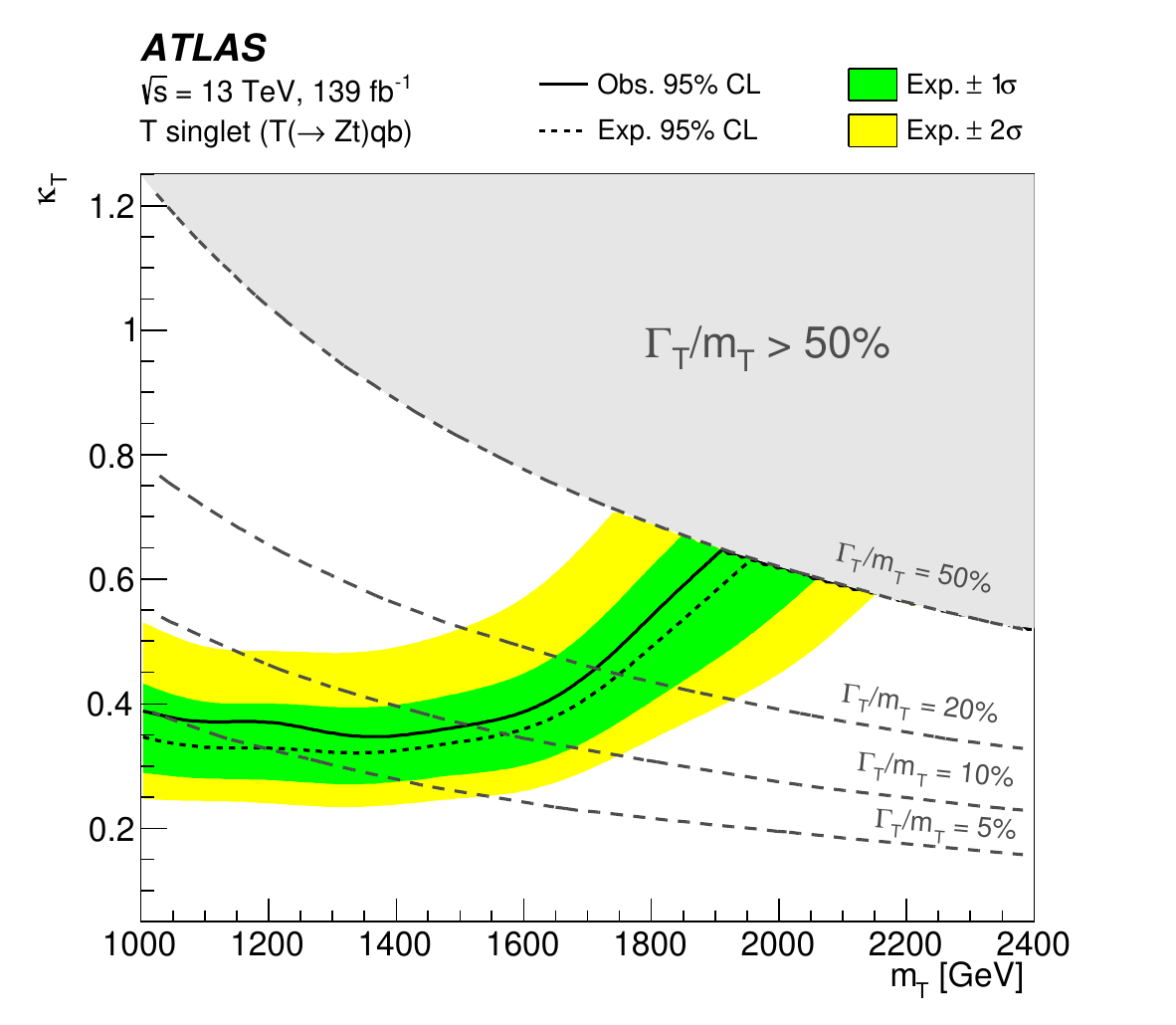}\label{fig:results_data_vlq_map2d}}
\subfloat[]{ \includegraphics[width=0.45\textwidth]{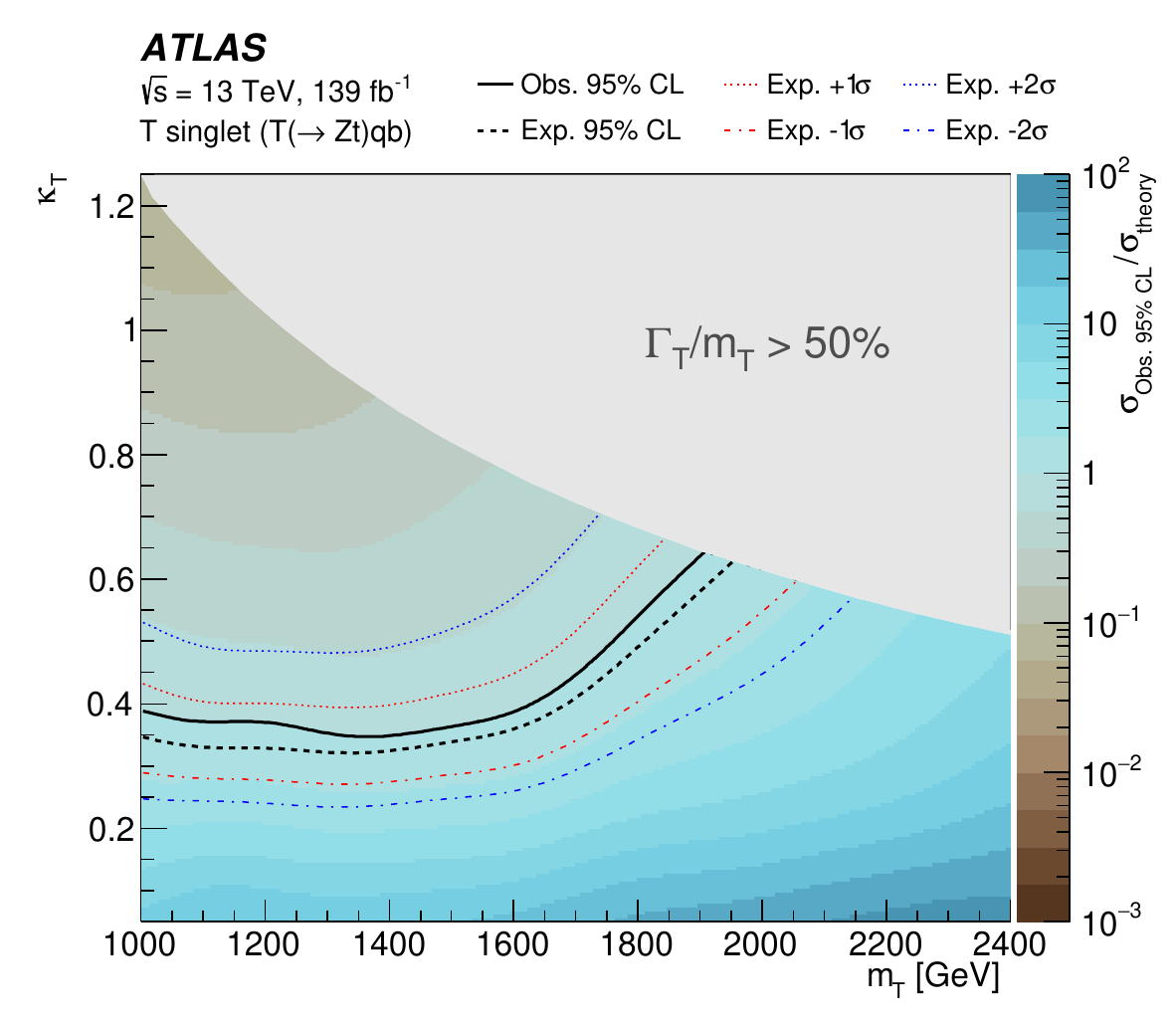}\label{fig:results_data_vlq_map2d_MERGED}}
\caption{Exclusion limits in terms of the universal coupling
constant \kappat as a function of the $T$ quark mass in the singlet SU(2) scenario, in the regime
where $\Gamma_T/m_T\leq50\%$, for which the theory calculations are known to be valid.
(a) Expected (dashed line) and observed (solid line) 95\% CL exclusion limits on \kappat as a function of the $T$ quark
mass. Different $\Gamma_T/m_T$ hypotheses are shown as dashed lines. The shaded bands correspond to $\pm1$ and $\pm2$
standard deviations around the expected limit.
(b) Observed 95\% CL upper limits on the $T$ quark signal cross-section divided by the theoretical value as a function of \kappat and the $T$ quark mass. The observed (expected) 95\% CL exclusion limits on the parameters are drawn as solid (dashed) lines, with all values of \kappat above the contour line being excluded at each mass point. The $\pm1$ and $\pm2$ standard deviations around the expected limit are also shown.}
\label{fig:results_data_vlq_maps}
\end{figure}

The two-dimensional exclusion contours in the plane $(m_T,\kappat)$ of the single vector-like $T$
quark production, shown in Figure~\ref{fig:results_data_vlq_maps}, were similarly obtained
using the internal weights of the generated signal samples. Figure \ref{fig:results_data_vlq_map2d}
provides exclusion limits on \kappat and the $T$ quark mass, while Figure~\ref{fig:results_data_vlq_map2d_MERGED}
shows the exclusion limits on the signal cross-section normalised to the theoretical value, derived also as a function of the mass and coupling.
Masses of the vector-like $T$ quark below $1.6$~\TeV are excluded for a value of the coupling
\kappat$>0.4$. For \kappat$>0.5$, the model is mostly excluded for masses $m_T<1.8$~\TeV.

\begin{figure}[t]
\centering{}
\includegraphics[width=0.45\textwidth]{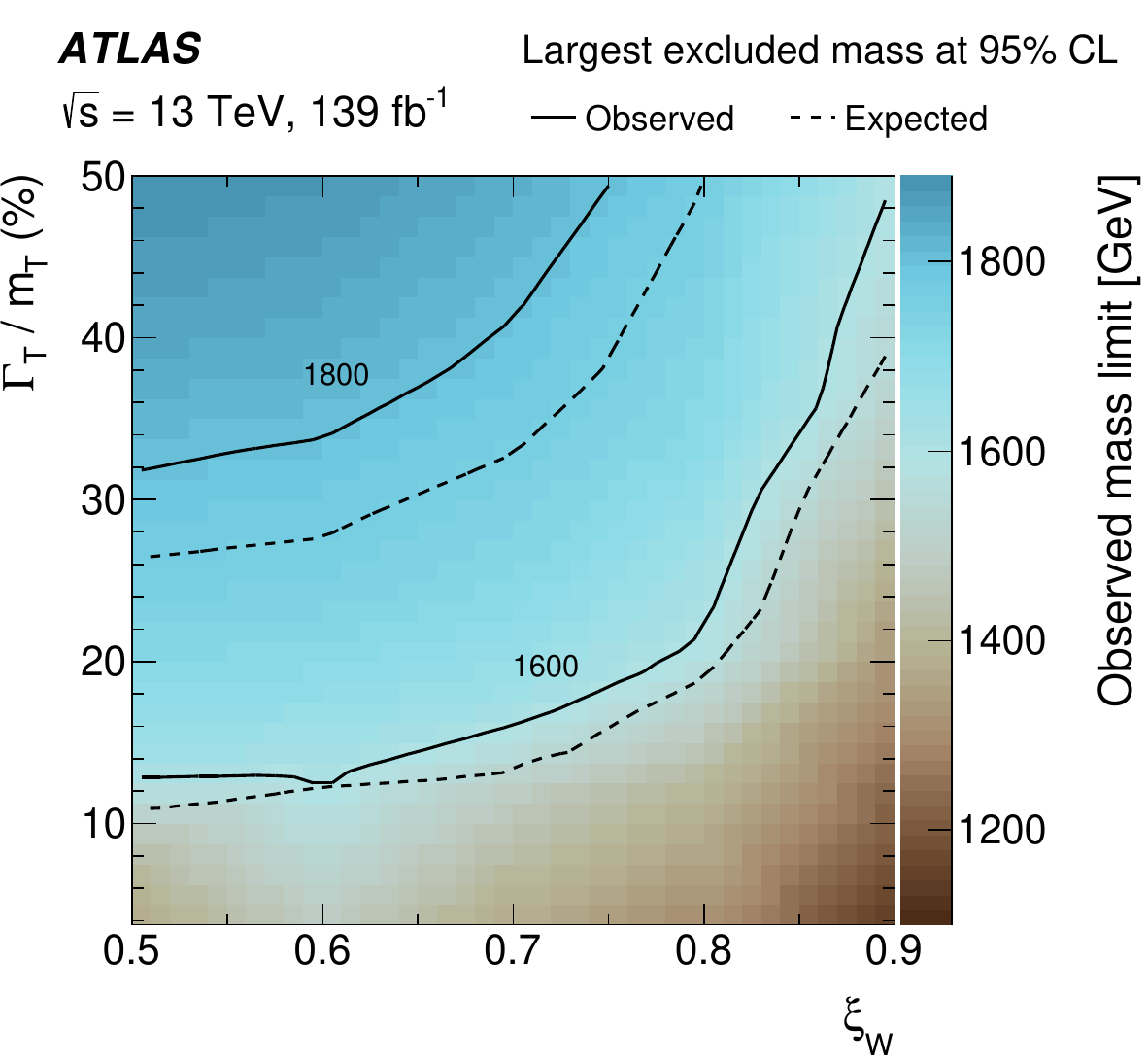}
\caption{Observed 95\% CL exclusion limits on the $T$ quark mass in the singlet SU(2) scenario as a function of the relative resonance width $\Gamma_T/m_T$ and the relative coupling parameter $\xi_W$. The solid (dashed) contour lines indicate observed (expected) exclusion limits of equal mass in units of \GeV.}
\label{fig:results_data_vlq_map2d_BR}
\end{figure}

The exclusion limits on the $T$ quark mass can be given in a more generalised representation of the
parameter space. Figure~\ref{fig:results_data_vlq_map2d_BR} displays the largest excluded mass as a function of
the relative resonance width $\Gamma_T/m_T$ and the relative coupling parameter $\xi_W$, which, as discussed in Section~\ref{sec:signal}, controls the
branching ratio $\mathcal{B}(T\to Wb)$ and fully contrains $\mathcal{B}(T\to tZ)$.
As expected, the exclusion limits on the $T$ quark mass are degraded for larger values of $\xi_W$, corresponding to lower values of the branching ratio associated with the decay explored in this analysis, $\mathcal{B}(T\rightarrow Zt)$.

%
%
%
%
%
%
%
%
%
%
%
%
%
%


%

\FloatBarrier

\section{Conclusions}
\label{sec:conclusions}

This paper reports the search for anomalous production of events with a single top quark and large missing transverse momentum using the $139$~fb$^{-1}$ dataset of $pp$ collisions at $\sqrt{s} = 13$~\TeV collected by ATLAS at the LHC during the Run~2 data-taking.

No deviations from the SM predictions are observed and the results are interpreted in terms of upper limits on the cross-sections of BSM models expected to populate the considered phase space. The models are the production of DM in association with a single top quark via a scalar or vector mediator, and the production of a single vector-like $T$ quark in the singlet scenario decaying into $tZ(\rightarrow\nu\bar{\nu})$. The observed limits on cross-sections are used to exclude ranges of values of model parameters: the scalar (vector) mediated production of DM in association with a single top quark is excluded for masses of the scalar (vector) particle up to $4.3$ ($2.3$)~\TeV for $m_\chi=1$~\GeV and couplings $\lambda_q=0.6$ and $y_\chi=0.4$ ($a=0.5$ and $g_\chi=1$). This result improves the limits on the mediator mass of the scalar (vector) model by approximately $800$ ($300$)~\GeV when compared to the previous analysis of \emph{mono-top} final states.
The production of a single vector-like $T$ quark decaying into $tZ(\rightarrow\nu\bar{\nu})$ is excluded for masses of the singlet $T$ quark up to $1.8$~\TeV, assuming $\kappa_T=0.5$. This limit outperforms previous results using the \emph{mono-top} topology by approximately $400$~\GeV.
This significant improvement partially comes from the refined object reconstruction and calibration, as well as the analysis strategy, optimising the event preselection and applying a \xgb algorithm to improve the background discrimination. The increased dataset statistics also play a significant role in the improved sensitivity.
%


%

\FloatBarrier

\section*{Acknowledgements}
%

%
%

%
%

We thank CERN for the very successful operation of the LHC and its injectors, as well as the support staff at
CERN and at our institutions worldwide without whom ATLAS could not be operated efficiently.

The crucial computing support from all WLCG partners is acknowledged gratefully, in particular from CERN, the ATLAS Tier-1 facilities at TRIUMF/SFU (Canada), NDGF (Denmark, Norway, Sweden), CC-IN2P3 (France), KIT/GridKA (Germany), INFN-CNAF (Italy), NL-T1 (Netherlands), PIC (Spain), RAL (UK) and BNL (USA), the Tier-2 facilities worldwide and large non-WLCG resource providers. Major contributors of computing resources are listed in Ref.~\cite{ATL-SOFT-PUB-2023-001}.

We gratefully acknowledge the support of ANPCyT, Argentina; YerPhI, Armenia; ARC, Australia; BMWFW and FWF, Austria; ANAS, Azerbaijan; CNPq and FAPESP, Brazil; NSERC, NRC and CFI, Canada; CERN; ANID, Chile; CAS, MOST and NSFC, China; Minciencias, Colombia; MEYS CR, Czech Republic; DNRF and DNSRC, Denmark; IN2P3-CNRS and CEA-DRF/IRFU, France; SRNSFG, Georgia; BMBF, HGF and MPG, Germany; GSRI, Greece; RGC and Hong Kong SAR, China; ISF and Benoziyo Center, Israel; INFN, Italy; MEXT and JSPS, Japan; CNRST, Morocco; NWO, Netherlands; RCN, Norway; MEiN, Poland; FCT, Portugal; MNE/IFA, Romania; MESTD, Serbia; MSSR, Slovakia; ARRS and MIZ\v{S}, Slovenia; DSI/NRF, South Africa; MICINN, Spain; SRC and Wallenberg Foundation, Sweden; SERI, SNSF and Cantons of Bern and Geneva, Switzerland; MOST, Taipei; TENMAK, T\"urkiye; STFC, United Kingdom; DOE and NSF, United States of America.

Individual groups and members have received support from BCKDF, CANARIE, CRC and DRAC, Canada; CERN-CZ, PRIMUS 21/SCI/017 and UNCE SCI/013, Czech Republic; COST, ERC, ERDF, Horizon 2020, ICSC-NextGenerationEU and Marie Sk{\l}odowska-Curie Actions, European Union; Investissements d'Avenir Labex, Investissements d'Avenir Idex and ANR, France; DFG and AvH Foundation, Germany; Herakleitos, Thales and Aristeia programmes co-financed by EU-ESF and the Greek NSRF, Greece; BSF-NSF and MINERVA, Israel; Norwegian Financial Mechanism 2014-2021, Norway; NCN and NAWA, Poland; La Caixa Banking Foundation, CERCA Programme Generalitat de Catalunya and PROMETEO and GenT Programmes Generalitat Valenciana, Spain; G\"{o}ran Gustafssons Stiftelse, Sweden; The Royal Society and Leverhulme Trust, United Kingdom.

In addition, individual members wish to acknowledge support from Chile: Agencia Nacional de Investigaci\'on y Desarrollo (FONDECYT 1190886, FONDECYT 1210400, FONDECYT 1230987); China: National Natural Science Foundation of China (NSFC - 12175119, NSFC 12275265, NSFC-12075060); European Union: European Research Council (ERC - 948254), Horizon 2020 Framework Programme (MUCCA - CHIST-ERA-19-XAI-00), Italian Center for High Performance Computing, Big Data and Quantum Computing (ICSC, NextGenerationEU), Marie Sklodowska-Curie Actions (EU H2020 MSC IF GRANT NO 101033496); France: Agence Nationale de la Recherche (ANR-20-CE31-0013, ANR-21-CE31-0013, ANR-21-CE31-0022), Investissements d'Avenir Idex (ANR-11-LABX-0012), Investissements d'Avenir Labex (ANR-11-LABX-0012); Germany: Baden-Württemberg Stiftung (BW Stiftung-Postdoc Eliteprogramme), Deutsche Forschungsgemeinschaft (DFG - 469666862, DFG - CR 312/5-1); Italy: Istituto Nazionale di Fisica Nucleare (FELLINI G.A. n. 754496, ICSC, NextGenerationEU); Japan: Japan Society for the Promotion of Science (JSPS KAKENHI 22H01227, JSPS KAKENHI 22KK0227, JSPS KAKENHI JP21H05085, JSPS KAKENHI JP22H04944); Netherlands: Netherlands Organisation for Scientific Research (NWO Veni 2020 - VI.Veni.202.179); Norway: Research Council of Norway (RCN-314472); Poland: Polish National Agency for Academic Exchange (PPN/PPO/2020/1/00002/U/00001), Polish National Science Centre (NCN 2021/42/E/ST2/00350, NCN OPUS nr 2022/47/B/ST2/03059, NCN UMO-2019/34/E/ST2/00393, UMO-2020/37/B/ST2/01043, UMO-2021/40/C/ST2/00187); Slovenia: Slovenian Research Agency (ARIS grant J1-3010); Spain: BBVA Foundation (LEO22-1-603), Generalitat Valenciana (Artemisa, FEDER, IDIFEDER/2018/048), La Caixa Banking Foundation (LCF/BQ/PI20/11760025), Ministry of Science and Innovation (RYC2019-028510-I, RYC2020-030254-I), PROMETEO and GenT Programmes Generalitat Valenciana (CIDEGENT/2019/023, CIDEGENT/2019/027); Sweden: Swedish Research Council (VR 2018-00482, VR 2022-03845, VR 2022-04683), Knut and Alice Wallenberg Foundation (KAW 2017.0100, KAW 2018.0157, KAW 2018.0458, KAW 2019.0447); Switzerland: Swiss National Science Foundation (SNSF - PCEFP2\_194658); United Kingdom: Leverhulme Trust (Leverhulme Trust RPG-2020-004); United States of America: Neubauer Family Foundation.

%
%


%
%
%
%
%
%
%

%
%
%
%
%
%

%
%

%
%
%
\printbibliography
\clearpage
\input{atlas_authlist}
%

%
%

%

%
%
%
%

\end{document}

%% file: atlas_authlist.tex
 
\begin{flushleft}
\hypersetup{urlcolor=black}
{\Large The ATLAS Collaboration}

\bigskip

\AtlasOrcid[0000-0002-6665-4934]{G.~Aad}$^\textrm{\scriptsize 102}$,
\AtlasOrcid[0000-0002-5888-2734]{B.~Abbott}$^\textrm{\scriptsize 120}$,
\AtlasOrcid[0000-0002-1002-1652]{K.~Abeling}$^\textrm{\scriptsize 55}$,
\AtlasOrcid[0000-0001-5763-2760]{N.J.~Abicht}$^\textrm{\scriptsize 49}$,
\AtlasOrcid[0000-0002-8496-9294]{S.H.~Abidi}$^\textrm{\scriptsize 29}$,
\AtlasOrcid[0000-0002-9987-2292]{A.~Aboulhorma}$^\textrm{\scriptsize 35e}$,
\AtlasOrcid[0000-0001-5329-6640]{H.~Abramowicz}$^\textrm{\scriptsize 151}$,
\AtlasOrcid[0000-0002-1599-2896]{H.~Abreu}$^\textrm{\scriptsize 150}$,
\AtlasOrcid[0000-0003-0403-3697]{Y.~Abulaiti}$^\textrm{\scriptsize 117}$,
\AtlasOrcid[0000-0002-8588-9157]{B.S.~Acharya}$^\textrm{\scriptsize 69a,69b,m}$
% ...
\bigskip
\\

% ...
$^{29}$Physics Department, Brookhaven National Laboratory, Upton NY; United States of America.\\
% ...
$^{35}$$^{(a)}$Facult\'e des Sciences Ain Chock, R\'eseau Universitaire de Physique des Hautes Energies - Universit\'e Hassan II, Casablanca;$^{(b)}$Facult\'{e} des Sciences, Universit\'{e} Ibn-Tofail, K\'{e}nitra;$^{(c)}$Facult\'e des Sciences Semlalia, Universit\'e Cadi Ayyad, LPHEA-Marrakech;$^{(d)}$LPMR, Facult\'e des Sciences, Universit\'e Mohamed Premier, Oujda;$^{(e)}$Facult\'e des sciences, Universit\'e Mohammed V, Rabat;$^{(f)}$Institute of Applied Physics, Mohammed VI Polytechnic University, Ben Guerir; Morocco.\\
% ...
$^{49}$Fakult\"{a}t Physik , Technische Universit{\"a}t Dortmund, Dortmund; Germany.\\
% ...
$^{55}$II. Physikalisches Institut, Georg-August-Universit\"{a}t G\"ottingen, G\"ottingen; Germany.\\
% ...
$^{69}$$^{(a)}$INFN Gruppo Collegato di Udine, Sezione di Trieste, Udine;$^{(b)}$ICTP, Trieste;$^{(c)}$Dipartimento Politecnico di Ingegneria e Architettura, Universit\`a di Udine, Udine; Italy.\\
% ...
$^{102}$CPPM, Aix-Marseille Universit\'e, CNRS/IN2P3, Marseille; France.\\
% ...
$^{117}$Department of Physics, New York University, New York NY; United States of America.\\
% ...
$^{120}$Homer L. Dodge Department of Physics and Astronomy, University of Oklahoma, Norman OK; United States of America.\\
% ...
$^{150}$Department of Physics, Technion, Israel Institute of Technology, Haifa; Israel.\\
$^{151}$Raymond and Beverly Sackler School of Physics and Astronomy, Tel Aviv University, Tel Aviv; Israel.\\
% ...
$^{a}$ Also Affiliated with an institute covered by a cooperation agreement with CERN.\\
$^{b}$ Also at An-Najah National University, Nablus; Palestine.\\
$^{c}$ Also at Borough of Manhattan Community College, City University of New York, New York NY; United States of America.\\
$^{d}$ Also at Center for High Energy Physics, Peking University; China.\\
$^{e}$ Also at Center for Interdisciplinary Research and Innovation (CIRI-AUTH), Thessaloniki; Greece.\\
$^{f}$ Also at Centro Studi e Ricerche Enrico Fermi; Italy.\\
% ...
$^{m}$ Also at Department of Physics, King's College London, London; United Kingdom.\\
% ...

\end{flushleft}
